\documentstyle[aps,subeqn,multicol,preprint]{revtex}

\def\biss{\prime\prime}
\def\lessgtr{\raise2.5pt\hbox{$<$}\llap{\lower2.5pt\hbox{$>$}}}
\def\gtrless{\raise2.5pt\hbox{$>$}\llap{\lower2.5pt\hbox{$<$}}}

\begin{document}

\title{The evolution of vibrational excitations in glassy systems }
\author{W.~G{\"o}tze and M.R. Mayr \\
Physik--Department, Technische Universit{\"a}t M{\"u}nchen,
 85747
Garching, Germany}

\maketitle

\vspace{0.5 truecm}

\begin{abstract}

The equations of the mode--coupling theory (MCT) for ideal
liquid--glass transitions are used for a discussion of the
evolution of the density--fluctuation spectra of glass--forming
systems for frequencies within the dynamical window between the
band of high--frequency motion and the band of
low--frequency--structural--relaxation processes. It is shown that
the strong interaction between density fluctuations with
microscopic wave length and the arrested glass structure causes an
anomalous--oscillation peak, which exhibits the properties of the
so--called boson peak. It produces an elastic modulus which
governs the hybridization of density fluctuations of mesoscopic
wave length with the boson--peak oscillations. This leads to the
existence of high--frequency sound with properties as found by
X--ray--scattering spectroscopy of glasses and glassy liquids. The 
results of the theory are demonstrated for a model of the
hard--sphere system. It is also derived that certain schematic MCT
models, whose spectra for the stiff--glass states can be expressed
by elementary formulas, provide reasonable approximations for the
solutions of the general MCT equations.

\bigskip

\end{abstract}

\section{Introduction}

The excitation spectra of normal liquids are located in the THz
band. But if the liquid can be supercooled new spectra evolve
within the GHz band as precursors of the glass transition. These
spectra are due to structural--relaxation processes. In addition
there appear vibrational spectra for frequencies $\omega$ above
0.1 THz which are characteristic for glasses and glass--forming
liquids, namely high--frequency sound and the so--called boson
peak. In this paper a microscopic theory shall be presented for
these vibrational excitations.

High--frequency sound was discovered only recently by
X--ray--scattering spectroscopy in various glass--forming systems:
in aqueous solutions of LiCl \cite{Masciovecchio96}, glycerol
\cite{Masciovecchio96,Masciovecchio98}, silica
\cite{Foret96,Benassi96,Masciovecchio97,Rat99}, orthoterphenyl
\cite{Masciovecchio98,Monaco98}, 0.4 Ca(NO$_3)_2$ 0.6 KNO$_3$
\cite{Matic99}, and B$_2$O$_3$ \cite{Matic99}. Molecular--dynamics
simulations identified these modes for models of argon glass
\cite{Mazzacurati96,Sampoli98}, ZnCl$_2$ \cite{Ribeiro98}, and
silica \cite{Dellanna98,Taraskin99b}. Five features of
high--frequency sound can be inferred from the cited work. First,
the resonance position $\Omega_q^{\max}$ of the dynamical
structure factor $S_q (\omega)$ is a linear function of the wave
number $q, \Omega_q^{\max}= v_\infty q$, and this for $q$ as large
as 4 to 8 $nm^{-1}$. Here $v_\infty$ is the same high--frequency
sound speed as known from Brillouin--scattering spectroscopy done
for wave vectors which are about two orders of magnitude smaller.
Second, the resonance width $\Gamma_q$ exhibits a quadratic
wave--vector dependence: $\Gamma_q = \gamma q^2$. Hence density
fluctuations propagate as known from the theory of elastic media
for wave lengths down to the order of the inter--particle
distances \cite{Masciovecchio96}. This holds with the reservation
that the spectrometer resolution of about 0.4 THz causes
considerable uncertainties in the data for $\Omega_q^{\max}$ and
$\Gamma_q$ and that it is difficult to separate the
inelastic--scattering signal from the huge background due to the
resolution--broadened quasi--elastic scattering. The damping of
ordinary sound in crystals or liquids depends strongly on
temperature. As a third remarkable feature of high--frequency
sound one observes that $\Gamma_q$ exhibits only a weak
temperature dependence. For silica and orthoterphenyl no
temperature dependence of $\Gamma_q$ could be detected even though
$\Omega_q^{\max}$ decreases with increasing temperature
considerably
\cite{Masciovecchio98,Benassi96,Masciovecchio97,Monaco98}. For
hydrodynamic sound the resonances exhaust the inelastic spectrum
of the density fluctuations. Contrary to this one infers from the
simulation work \cite{Mazzacurati96,Sampoli98,Taraskin99b} as a
fourth property of high--frequency sound that the resonances are
superimposed on a broad intense background spectrum of $S_q
(\omega)$. As a fifth property one finds for the simulation
results of silica \cite{Taraskin99b} that the background spectrum
exhibits a low--frequency threshold. This conclusion is supported
by the measurement of the X--ray--scattering spectrum of densified
silica \cite{Rat99}, which demonstrates that there is a
suppression of $S_q (\omega)$ for small $\omega$ relative to the
hydrodynamics description for the dynamical structure factor.

The so--called boson peak has been known for quite some time since
it can be detected in the spectra obtained by standard techniques.
It was measured for all the above cited systems, for example by
Raman scattering for LiCl \cite{Tao91b} and by neutron scattering
for silica \cite{Foret96,Buchenau86}. It was found by
molecular--dynamics simulations for ZnCl$_2$ \cite{Foley95} and
silica \cite{Taraskin99,Horbach99b}. Six features are typical for
these peaks. First of all, the peaks are due to soft excitations.
The position $\omega_P$ of the peak maximum is several times
smaller than the Debye frequency $\omega_D$ of the system. In
silica the peak causes an enhancement of the so--called density of
states by a factor of about seven relative to the Debye spectrum
\cite{Buchenau86}. Second, the anomalous peak is due to
quasi--harmonic oscillations. Originally, this was concluded
indirectly, since this assumption can explain the enhancement of
the specific heat above Debye's $T^3$--law for temperatures $T$
near 10K \cite{Buchenau86}. Normal--mode analysis for a model of
silica showed this result explicitly \cite{Taraskin99}. Third, the
peak is skewed. There seems to be a low--frequency threshold
$\Omega_-$ so that the high--frequency wing of the peak extends
further than the one on the low--frequency side
\cite{Foret96,Tao91b,Horbach99b}. Fourth, the peak appears to be
related to some instability of the system since $\omega_P$ and
$\Omega_-$ decrease to zero upon heating the system towards some
characteristic temperature $T_{bp}$ \cite{Tao91b,Horbach99b}. The
mentioned four features are also exhibited by the density of
states of a harmonic lattice with a random distribution of force
constants \cite{Schirmacher98}. As a fifth property one observes
that a quasi--elastic relaxation peak appears for $T$ increasing
towards $T_{bp}$ which eventually buries the peak \cite{Tao91b}.
Sixth, the peak position $\omega_P$ of the dynamical structure
factor depends only weakly on wave vector $q$ if at all
\cite{Foret96,Horbach99b}.

The discussions of this paper are done within the framework of the
mode--coupling theory (MCT) for the evolution of glassy dynamics
in simple systems. This theory is based on closed
microscopic equations of motion for the density--correlation
functions \cite{Bengtzelius84,Leutheusser84}. Previous work done in this context
focused on elucidating the properties of the
structural--relaxation phenomena, as described in Refs.
\cite{Goetze91b,Goetze92} and the papers quoted there. In the
following it will be shown that MCT also implies a theory for the
evolution of anomalous--oscillation peaks (AOP) with the same six
properties specified in the preceding paragraph for the so--called
boson peak. Furthermore it will be demonstrated that the AOP
manifests itself in resonances of the density--fluctuation spectra
for wave--vectors up to half the value of the Debye vector $q_D$,
which exhibit the five features listed above for high--frequency
sound.

MCT deals with states of matter for which the structure factor
$S_q$ depends smoothly on wave vector $q$ and on control
parameters like the temperature $T$ and density $\rho$. The
equations of motion of the basic version of this theory, which
will be used in this paper, exhibit a bifurcation from ergodic
liquid dynamics to non--ergodic glass dynamics if $T$ or $\rho$
cross critical values $T_c$ or $\rho_c$ respectively. This
bifurcation provides a model for an ideal liquid--glass
transition. In the extended version of MCT the singular transition
is replaced by regular crossover \cite{Goetze92}. The crossover is
connected with the evolution of structural relaxation. Extensive
tests of the MCT predictions have been performed in recent years.
Let us only mention one group of such tests which are of
particular relevance for the following discussions. The MCT
results for the Debye--Waller factors $f_q$ of the glass depend
non trivially on $q$ and on $T$ or $\rho$. The results calculated
for the hard--sphere system \cite{Bengtzelius84} agree with the
data measured for hard--sphere colloids \cite{Megen94b}.
Similarly, the $f_q$ calculated for a Lennard--Jones mixture
\cite{Nauroth97} agree within 10\% with the values deduced from
molecular--dynamics studies \cite{Gleim98}. These assessments of
the theory and many other tests, which are reviewed in Ref.
\cite{Goetze99}, show that MCT describes properly some essential
features of structural relaxation. Since the MCT equations are
rather involved most of the original work was done first for
schematic models. These are truncations of the large set of
equations to sets dealing with a few correlators only. The
truncations are done with the intention to reduce the complexity
of the mathematical machinery without loosing the essence of the
features to be analyzed for the solution. It was noticed for
schematic models dealing with a single correlator that the spectra
for the states for $T \ll T_c$ or $\rho \gg \rho_c$ exhibit a
broad peak, which is due to a superposition of harmonic oscillator
spectra \cite{Goetze91b,Goetze84}. Tao et al. recognized that the
evolution of these peaks due to changes of control parameters was
similar to what they had measured for the boson peak of glassy
aqueous LiCl \cite{Tao91b}. Therefore they concluded that MCT
implies a theory for the boson--peak spectrum. This conclusion was
corroborated by a series of studies, where solutions of schematic
models were used to describe quantitatively spectra of glassy
liquids
\cite{Franosch97a,Krakoviack97,Singh98,Ruffle97,Ruffle98,Ruffle99}.
The fits described structural--relaxation spectra in windows of
several orders of magnitude in size in addition to parts of the
boson--peak spectra for high frequencies. In the following this
earlier work will be extended to a systematic theory for the
general MCT equations. 
This theory implies, in particular, the suggestion that the 
characteristic temperature $T_{bp}$ for the boson-peak dynamics is 
identical with $T_c$. Our results will be demonstrated
comprehensively for a model of the hard--sphere system (HSS).

The paper is organized as follows. In Sec.II the basic equations
for our calculations are formulated and the details for the HSS
work are specified. The evolution of structural relaxation will be
demonstrated in order to put the discussion in the proper context
with the theory of the glass transition. Then, in Sec.III, the
results for the evolution of the anomalous--oscillation peak and
for high--frequency sound are presented. These oscillation
features of the MCT dynamics are described within a
generalized--hydrodynamics approach. In Sec.IV the phenomena are
explained within a schematic--model analysis. In the concluding
Sec.V our findings are discussed.

\section{Basic Equations}
\subsection{\bf{A mode--coupling--theory model for a dense simple system}}

The basic quantity specifying the equilibrium structure of a
simple system is the structure factor $S_q = \langle \mid
\rho_{\vec q}\mid^2 \rangle$, which is the canonical average of
the squared modulus of the density fluctuations of wave vector
$\vec q : \rho_{\vec q} = \sum_\alpha \exp (i \vec q \vec
r_\alpha) / \sqrt{N}$. Here $\alpha$ labels the N particles at
positions ${\vec r}_\alpha$ in the system of density $\rho$. The
structure factor of amorphous systems depends on the wave vector
modulus $q = \mid \vec q \mid$ only and it is usually expressed in
terms of the direct correlation function $c_q : S_q = 1/ (1- \rho
c_q)$ \cite{Hansen86}. The most relevant variables describing the
dynamics of structure changes as function of time $t$ are the
density correlators $\phi_q (t) = \langle \rho_{\vec q}^*
(t)\rho_{\vec q} \rangle / S_q$. The short time asymptote of these
functions is specified by a characteristic frequency $\Omega_q :
\phi_q (t) = 1 - \frac{1}{2} (\Omega_q t)^2 + \cdots$. In the
small--wave--vector limit one gets the dispersion law for
sound; $\Omega_q = v_0 q + O (q^3)$, where $v_0$
denotes the isothermal sound speed. For general wave vectors one
obtains $\Omega_q^2 = v^2 q^2 / S_q$, where $v$ denotes the
thermal velocity of the particles \cite{Hansen86}.
In particular $v_0^2 = v^2 / S_0$. Within the Zwanzig--Mori
formalism one can derive the exact equation of motion
\begin{mathletters}
\label{Gl1}
\begin{equation}
\label{Gl1a}
\ddot{\phi}_q (t) + \Omega_q^2 \phi_q (t) + \Omega_q^2 \int_0^t
m_q (t-t^\prime) \dot{\phi}_q (t^\prime)dt^\prime = 0 \,\, .
\end{equation}
The relaxation kernel $m_q (t)$ is a correlation function of
fluctuating forces \cite{Hansen86}. Let us introduce here and in
the following Fourier--Laplace transformations to map $\phi_q,
m_q$ and similar functions from the time domain onto the frequency
domain according to the convention $\phi_q (\omega) = i
\int_0^\infty \exp (i\omega t) \phi_q (t) dt = \phi_q^\prime
(\omega) + i \phi_q^{\prime\prime} (\omega)$. The reactive part
$\phi_q^\prime (\omega)$ and the dissipative part or the spectrum
$\phi_q^{\prime\prime} (\omega)$ are connected via a
Kramers--Kronig relation. Equation (\ref{Gl1a}) is equivalent to the
representation $\phi_q (\omega) = - 1 / [\omega -
\Omega_q^2 / [\omega + \Omega_q^2 m_q (\omega)]]$. The fluctuation
dissipation theorem connects $\phi_q (\omega)$ with the dynamical
susceptibility $\chi_q (\omega) : \chi_q (\omega) = [\omega\phi_q
(\omega) + 1] \chi_q^T$. Here $ \chi_q^T = S_q / (\rho \mu v^2)$,
with $\mu$ denoting the mass of the particles, is the isothermal 
compressibility
\cite{Hansen86}. Therefore Eq. (\ref{Gl1a}) is equivalent to
\begin{equation}
\label{Gl1b}
\chi_q (\omega) / \chi_q^T = -\Omega_q^2 /\left[\omega^2 -
\Omega_q^2 + \Omega_q^2 \omega m_q (\omega)\right] \,\, .
\end{equation}
\end{mathletters}

Within the MCT kernel $m_q(t)$ is written as the sum of a regular
contribution and a
mode--coupling contribution, describing the cage effect of dense
systems within Kawasaki's factorization approximation. In this
paper we neglect the regular term. Hence
one gets \cite{Bengtzelius84}:
\begin{mathletters}
\label{Gl2}
\begin{equation}
\label{Gl2a}
m_q (t) = {\cal{F}}_q \left[ \phi (t)\right] \,\, ,
\end{equation}
where the mode--coupling functional  ${\cal {F}}_q$, considered as
functional of the dummy variable $\tilde f$, is:
${\cal{F}}_q [\tilde f] = \sum_{\vec{k}+\vec{p}=\vec{q}}
V(\vec{q},\vec{k} \vec{p}) \tilde f_k \tilde{f}_p$. The coupling
coefficients $V\vec{(q}, \vec{k} \vec{p})$ are determined by the
equilibrium structure:
\begin{equation}
\label{Gl2b}
V\vec{(q}, \vec{k} \vec{p}) = \rho S_q S_k S_p \left[\vec q (\vec
k c_k + \vec p c_p) \right]^2 / (2q^4) \,\, .
\end{equation}
Let us approximate wave--vector integrals by Riemann sums,
obtained by choosing an equally--spaced wave--vector grid of M
terms. Thus the wave--vector index $q$ will be understood as a
label running from 1 to M. Correlators $\phi (t)$, kernels etc.
are considered as M-component vectors. After this discretization
the functional is a quadratic polynomial
\begin{equation}
\label{Gl2c}
{\cal{F}}_q [\tilde f] = \sum_{kp=1}^M V_{q,kp} \tilde f_k \tilde
f_p \,\, .
\end{equation}
\end{mathletters}
The positive coefficients $V_{q,kp}$ are related trivially to
$V(\vec{q},\vec{k} \vec{p})$ in Eq. (\ref{Gl2b}); for details the 
reader is referred to Ref. \cite{Franosch97}.

Equations (\ref{Gl1}) and (\ref{Gl2}) are closed. They define a unique 
solution
for all parameters $\Omega_q > 0, V_{q,kp} \geq 0$. The solution
has all the properties of a correlator, i.e. the $\phi_q (t)$ are
positive--definite functions given by
non--negative spectra $\phi_q^{\biss}(\omega)$. And the solutions
depend smoothly on $\Omega_q$ and $V_{q,kp}$ for all finite time
intervals \cite{Haussmann90,Goetze95b}. Thus, Eqs. (\ref{Gl1}) 
and (\ref{Gl2})
formulate a well--defined model for a dynamics. The model is
related to the physics of liquids by specifying $\Omega_q$ and
$V_{q,kp}$ in terms of particle interactions and the control
parameters $\rho$, $T$.

In the following the results will be demonstrated for the
hard--sphere system (HSS). The structure factor is independent of
T in this case, so that the only non--trivial control parameter is
the packing fraction $\varphi$ of the spheres of diameter $d :
\varphi = \pi d^3 \rho / 6$. The structure factor $S_q$ is
evaluated in the Percus--Yevick approximation \cite{Hansen86}.
Wave vectors will be considered up to a cutoff value $q^* = 40/d$
and they are discretized to $M = 300$ values. The sphere diameter
will be chosen as the unit of length, $d = 1$, and the unit of
time will be chosen such that the thermal velocity is $v = 2.5$.

\subsection{\bf{Ideal glass states}}

The specified model exhibits a fold bifurcation \cite{Goetze91b}.
For small coupling constants $V_{q,kp}$ the correlators $\phi_q
(t)$ and kernels  $m_q (t)$ decay sufficiently fast to zero for
times tending to infinity so that the spectra
$\phi_q^{\biss} (\omega)$ and $m_q^{\biss} (\omega)$ are
continuous in $\omega$. Density fluctuations, which are created at
time $t = 0$, disappear for long times; and the same holds for the
force fluctuations. The system approaches equilibrium for long
times as expected for an ergodic liquid. This implies
$\lim_{\omega \to 0} \omega m_q (\omega) = 0$ and one concludes
from Eq. (\ref{Gl1b}) that the static 
susceptibility $\hat \chi_q = \chi_q
(\omega \to 0)$ agrees with the thermodynamic susceptibility,
$\hat \chi_q = \chi_q^T$. For large coupling constants, on the
other hand, there is arrest of density fluctuations for long
times: $\phi_q (t \to \infty) = f_q$; $0 < f_q < 1$. Thus the
perturbed system does not return to the equilibrium state.
Similarly, there is arrest of the force fluctuations:
$m_q (t \to \infty) = C_q > 0$. The numbers $f_q, C_q$ are
connected by the equations \cite{Bengtzelius84}:
\begin{equation}
\label{Gl3}
f_q = C_q / (1 + C_q) \qquad , \qquad C_q = {\cal {F}}_q [f]
\qquad , \qquad q = 1, 2, \ldots , M \,\, .
\end{equation}
For this strong coupling solution the kernel exhibits a
zero--frequency pole, $\lim_{\omega \to 0} \omega m_q (\omega) = -
C_q$, and therefore one concludes from Eq. (\ref{Gl1b}) that the static
susceptibility is smaller than the thermodynamic one, $\hat\chi_q
< \chi_q^T$, since $\hat\chi_q = \chi_q^T / (1 + C_q)$. This is a
signature for a non--ergodic state \cite{Kubo57}. The system
reacts more stiffly than expected for a canonical averaging. The
dynamical structure factor $S_q (\omega) = S_q \phi_q^{\biss}
(\omega)$ exhibits a strictly elastic peak: $S_q (\omega) = \pi
S_q f_q \delta (\omega) +$ regular terms. This is the signature
for a solid with $f_q$ denoting its Debye--Waller factor. Hence, the
strong coupling solution deals with a disordered solid; it is a
model for an ideal glass state.
If one increases the coupling constants smoothly from small to
large values one finds a singular change of the solution from the
ergodic liquid to the non--ergodic glass state, i.e. an idealized
liquid--glass transition. For simple--liquid models the transition
occurs upon cooling at some critical temperature $T_c$ or upon
compression at some critical packing fraction $\varphi_c$
\cite{Bengtzelius84}. For the HSS model under study one finds
$\varphi_c \approx 0.516$ \cite{Franosch97}. If $\varphi$
increases above $\varphi^c$, the Debye--Waller factor $f_q$
increases above its value at the critical point, called the 
plateau $f_q^c$, 
as is demonstrated in Fig. \ref{fig_DWF}.

In the theory of crystalline solids one defines susceptibilities
with respect to the restricted ensemble of a given arrested
lattice. The MCT equations of motion allow for a similar
formulation of the equations of motion for the glass state
\cite{Goetze91b}. To see this, one has to map the density
correlators $\phi_q$ to new ones $\hat\phi_q$ by
\begin{mathletters}
\label{Gl4}
\begin{equation}
\label{Gl4a}
\phi_q (t) = f_q + (1 - f_q) \hat\phi_q (t) \,\, .
\end{equation}
If one introduces new characteristic frequencies $\hat\Omega_q$ by
\begin{equation}
\label{Gl4b}
\hat\Omega_q^2 = \Omega_q^2 / (1-f_q) \,\, ,
\end{equation}
one obtains the short time expansion $\hat\phi_q (t) = 1 -
\frac{1}{2} (\hat\Omega_q t)^2 + \cdots$ in analogy to what was
found for $\phi_q (t)$. Substitution of these results into Eq.
(\ref{Gl1a}) reproduces the MCT equations of motion with $\phi_q,
\Omega_q$ and $m_q$ replaced by $\hat\phi_q, \hat\Omega_q$ and
$\hat m_q$, respectively. Here the new relaxation kernels are
related to the original ones by
\begin{equation}
\label{Gl4c}
m_q (t) = C_q + (1 + C_q) \hat m_q (t) \,\, .
\end{equation}
For the dynamical susceptibility one obtains the formula $\chi_q
(\omega) = \hat\chi_q [1 + \omega \hat\phi_q (\omega)]$. The new
correlator has a vanishing long time limit, $\hat\phi_q (t \to
\infty) = 0$, the Fourier--Laplace transform exhibits the property
$\lim_{\omega \to 0} \omega \hat\phi_q (\omega) = 0$. Since the
equation of motion does not change its form, one gets as the
analogue of Eq. (\ref{Gl1b}) the expression of the susceptibility $\chi_q
(\omega)$ in terms of the polarization operator $\hat m_q
(\omega)$:
\begin{equation}
\label{Gl4d}
\chi_q (\omega) / \hat\chi_q = - \hat\Omega_q^2 / \left[ \omega^2
- \hat\Omega_q^2 +  \hat\Omega_q^2 \omega \hat m_q (\omega)\right]
\quad, \quad \hat \chi_q = \chi_q^T (1 - f_q)\,\, .
\end{equation}
\end{mathletters}
Combining Eq. (\ref{Gl3}) with Eq. (\ref{Gl4c}) one 
concludes $\hat m_q (t \to\infty) = 0$, i.e. the new kernel 
$\hat m_q (\omega)$ for the
amorphous solid exhibits a regular zero--frequency behaviour:
$\lim_{\omega \to 0} \omega \hat m_q (\omega) = 0$. 

The mentioned equation of motion for $\hat\phi_q (t)$ can be
closed by combining Eqs. (\ref{Gl4a}) and (\ref{Gl4c}) with 
Eqs. (\ref{Gl2a}) 
and (\ref{Gl2c}).
One finds an expression of the new kernel as a new mode--coupling
functional $\hat{\cal{F}}$ of the new correlators:
\begin{mathletters}
\label{Gl5}
\begin{equation}
\label{Gl5a}
\hat m_q (t) = \hat{\cal{F}}_q \left[ \hat\phi (t) \right] \,\, .
\end{equation}
$\hat {\cal{F}}$ is a sum of a linear term $\hat {\cal{F}}^{(1)}$
and a quadratic one  $\hat {\cal{F}}^{(2)}, \hat {\cal{F}}_q =
\hat {\cal{F}}_q^{(1)} + \hat {\cal{F}}_q^{(2)}$, where:
\begin{equation}
\label{Gl5b}
\hat  {\cal{F}}_q^{(1)} [\tilde f] = \sum_k \hat V_{q,k} \tilde
f_k \quad , \quad \hat V_{q,k} = 2(1-f_q) \sum_p V_{q, kp} f_p
(1-f_k) \,\, ,
\end{equation}
\begin{equation}
\label{Gl5c}
\hat {\cal{F}}_q^{(2)} [\tilde f] = \sum_{kp} \hat V_{q,kp} \tilde
f_k \tilde f_p \quad , \quad \hat V_{q, kp} = (1-f_q) V_{q, kp}
(1-f_k)(1 - f_p) \,\, .
\end{equation}
As a result equations of motion are produced, which are of the
same form as the MCT equations discussed in Sec.IIA.
But in addition to the quadratic mode--coupling term there appears
a linear one, and the values of the mode--coupling coefficients
from Eq. (\ref{Gl2b}) are renormalized to the coefficients $\hat V_{q,
kp}$.

The preceding Eqs. (\ref{Gl5}) are equivalent to equations relating the
density--fluctuation spectra $\hat \phi_q^{\biss} (\omega)$ with
the kernel spectra $\hat m_q^{\biss}(\omega) = \hat m_q^{(1)
\biss} (\omega) + \hat m_q^{(2) \biss} (\omega)$:
\begin{equation}
\label{Gl5d}
\hat m_q^{(1)\biss} (\omega) = \sum_k \hat V_{qk}
\hat\phi_k^{\biss} (\omega) \,\, ,
\end{equation}
\begin{equation}
\label{Gl5e}
\hat m_q^{(2) \biss}(\omega) = \frac{1}{\pi}\sum_{kp} \int
d\omega_1 \int d\omega_2 \hat V_{q, kp} \delta (\omega - \omega_1
- \omega_2) \hat \phi_k^{\biss} (\omega_1)\hat \phi_p^{\biss}
(\omega_2) \,\, .
\end{equation}
\end{mathletters}
One can interprete Eqs. (\ref{Gl4}) and (\ref{Gl5}) using the 
language of the
theory of boson fields, e.g. the phonon fields in crystals. $\hat
\chi_q (\omega) / \hat \chi_q$ is the field propagator and $\hat
\Omega_q$ is the bare--phonon--dispersion law. The kernel $\hat
m_q (\omega)$ is the phonon self energy.
Equations (\ref{Gl5d}) and (\ref{Gl5e}) are golden--rule 
expressions for the
phonon--decay rates. Kernel $\hat m^{(1)}$ describes elastic
scattering of the phonon from the disorder, produced by the
amorphous glass structure. Kernel $\hat m^{(2)}$ deals with the
decay of a phonon into two due to anharmonicities. The
glass structure influences the decay rates via the Debye--Waller
factors which enter Eqs. (\ref{Gl5b}) and (\ref{Gl5c}) for 
$\hat V_{q,k}$ and
$\hat V_{q,kp}$, respectively. The challenge is to evaluate this
probability $\hat m_q^{\biss} (\omega)$ self consistently;
the decay depends on the same phonons as one wants to
study.

\subsection{\bf {The glass--transition scenario}}

Figure \ref{fig_HSM_ISF_t} exhibits the evolution of the dynamics 
with increase of
the packing fraction $\varphi$. The wave
vector $q = 7.0$, used in the lower panel, is close to the
structure--factor--peak position where the static susceptibility
$\chi_q^T \propto S_q$ is high, while the wave vector $q = 3.4$,
used in the upper panel, deals with fluctuations where $\chi_q^T$
is very small (compare Fig. \ref{fig_DWF}). Figure \ref{fig_HSM_ISF_w} 
shows the equivalent
information for the fluctuation spectra $\phi_q^{\biss} (\omega)$.
In Ref. \cite{Franosch98} a further set of diagrams for
the wave vector $q = 10.6$ can be found.

The curves for $\varphi < \varphi_c$ with label $n = 1$ in 
Figs. \ref{fig_HSM_ISF_t}
and \ref{fig_HSM_ISF_w} refer to the packing fraction 
$\varphi = 0.276$. The
correlators exhibit strongly--damped oscillations; and the ideal
phonon resonances, to be expected for $m_q = 0$, are altered to
broadened bumps in the spectra. If the packing fraction increases
into the interval 0.9 $\varphi_c \leq \varphi < \varphi_c$, the
oscillatory features almost disappear for $t > 0.2$; and the shown
spectra $\phi_q^{\biss} (\omega)$ decrease monotonously with
increasing $\omega$. Simultaneously, the decay to equilibrium is
delayed to larger times. At the critical point the correlators
approach the plateau $f_q^c$ in a stretched manner, as shown by
the curves with label c in Fig. \ref{fig_HSM_ISF_t}. 
This process, which is called
critical decay, leads to a strong increase of the fluctuation
spectra with decreasing frequency, as shown in 
Fig. \ref{fig_HSM_ISF_w}. Increasing
$\varphi$ above the critical packing fraction $\varphi_c$ the
values for the long time limits $f_q$ increase. Since these limits
are approached exponentially fast for $\varphi \not= \varphi_c$
\cite{Goetze95b}, the correlation spectra become
$\omega$--independent for low frequencies. The value for this
white noise spectrum increases if $\varphi$ decreases towards
$\varphi_c$; this is a precursor phenomenon of the glass melting
at the transition point. If $\varphi$ is sufficiently far above
$\varphi_c$, one observes oscillations again, as can be seen for
the $\varphi = 0.6$ result in the upper panel of 
Fig. \ref{fig_HSM_ISF_t}. In this
case, the correlation spectra are not anymore monotone functions
of frequency. For $q = 3.4$, there occur two peaks for 
$\omega > 10$. The narrow peak is due to high--frequency 
phonon propagation. 
In addition there is an anomalous--oscillation peak (AOP)
for $\omega \approx 80$. For $q = 7$ a phonon peak is absent in
the spectrum for $\varphi = 0.6$ but an AOP is present, as is
shown in the lower panel of Fig. \ref{fig_HSM_ISF_w}.

The time scale for normal--state--liquid dynamics 
is set by the Debye frequency $\omega_D$. It is the same scale as
for the dynamics of the crystalline state of matter. For the
discussion of this normal condensed--matter dynamics it is
sufficient to consider a window of, say, two decades around $2 \pi
/ \omega_D $. This is demonstrated in Fig. \ref{fig_HSM_ISF_t} 
for the liquid state
with label $n = 1$ or for the glass state with label $n = 3$. If
the time increases from 0.01 to 1 the correlators $\phi_q (t)$
decrease from 0.9 to the long time limit. All
oscillations occur within this interval, which is also called the
transient regime. The spectra for these states are
located within a corresponding regime of microscopic excitations,
extending roughly between 0.3 and 300. For frequencies around and
below 0.3 there is only white noise for the specified normal
states. The glass transition is connected with a dynamics, called
glassy dynamics, which occurs for times longer and frequencies
smaller than the ones characterizing the transient. For reasons
which can be understood by asymptotic expansions about the
critical point \cite{Goetze91b}, the dynamics is stretched over
many decades. The glassy dynamics of the HSS is discussed
comprehensively in Refs. \cite{Franosch97,Fuchs98}. It is
impossible to view glassy dynamics adequately on linear scales for
$t$ or $\omega$. Conventionally, one represents the results on
logarithmic abscissas as done in Figs. \ref{fig_HSM_ISF_t} and 
\ref{fig_HSM_ISF_w}.
The bifurcation for $\varphi = \varphi_c$ or $T = T_c$ also
modifies the transient dynamics. The subject of the paper is the
study of these modifications.

\section{Evolution of the transient dynamics}
\subsection{\bf{Anomalous--oscillation peaks and high--frequency phonons}}

For a discussion of the transient motion it is sufficient to
consider dynamical windows of about two orders of magnitude. These
can be viewed more adequately on linear rather than on logarithmic
abscissas. There is no reason to consider
such fine tuning of control parameters relative to the critical
point as is necessary for a study of structural relaxation.
Therefore let us extract the relevant information for the
following discussion from Figs. \ref{fig_HSM_ISF_t} and \ref{fig_HSM_ISF_w} 
and replot it as Figs. \ref{fig_lin_ISF_t} and \ref{fig_lin_ISF_w}. 
Let us first consider the results for wave vector $q = 7$.
For the packing fraction $\varphi = 0.60$ the particles are
localized in such tight cages that the square root $\delta r =
\sqrt {\langle \delta r^2 (t \to \infty) \rangle}$ of the
long--time limit of the mean--squared displacement is only 5.0\%
of the particle diameter \cite{Fuchs98}. Thus one expects the
particles to bounce in their cages with an average frequency of
the order $\omega \approx 2\pi v/(4 \cdot \delta r) = 78$. This
explains qualitatively the oscillation around the equilibrium
value $f_{7.0}$ exhibited by $\phi_{7.0} (t)$ in 
Fig. \ref{fig_lin_ISF_t} for $t <
0.3$ and the corresponding "peak" of the spectrum in 
Fig. \ref{fig_lin_ISF_w}. The
maximum position $\omega_P \approx 75$ of this AOP is estimated
well by the crude formula. The AOP differs qualitatively from the
Lorentzian which one would expect for some damped harmonic
oscillation. The low frequency part of the peak decreases more
steeply with decreasing $\omega$ so that there appears some
threshold near $\omega = 40$. Below the threshold the spectrum is
flatter than expected for the wing of a Lorentzian. If $\varphi$
decreases to 0.567 the cages widen so that $\delta r = 0.070$
\cite{Fuchs98}. This explains the increase of the oscillation
period exhibited by curve $n = 3$ in Fig. \ref{fig_lin_ISF_t} 
and the corresponding
downward shift of $\omega_P$ in Fig. \ref{fig_lin_ISF_w}. 
The shift is accompanied
by a strong increase of the spectrum for $\omega \leq 10$. The
integral of the inelastic spectrum also increases, reflecting the
decrease of the elastic contribution $\pi f_q$ which is exhibited
in Fig. \ref{fig_DWF}. The described trends continue if $\varphi$ 
is decreased
further to 0.540 (curves $n = 4$). The threshold and the spectral
minimum for small frequencies are now replaced by a central peak
for $\omega \leq 20$. The maximum of the AOP, estimated from
$\delta r = 0.099$ \cite{Fuchs98} as $\omega_P \approx 40$, is 
buried under the tail of the central peak; it merely shows up as a
shoulder. The correlator still exhibits some small oscillation
before it reaches its long time limit $f_{7.0} = 0.96$ for $t >
0.2$, but it does not fall below $f_{7.0}$ anymore. At the
critical point $\delta r = \delta r^c = 0.183$ \cite{Fuchs98} and
so one estimates a position $\omega_P \approx 21$. But the
critical decay manifests itself by the appearance of a long time
tail of $\phi_{7.0} (t)$. The approach to the asymptote $f_{7.0}^c
= 0.85$ cannot be demonstrated on the linear time axis used in
Fig. \ref{fig_lin_ISF_t}. This tail leads to a strong enhancement 
of the central
peak, so that the AOP cannot be identified. This
trend continues if a packing fraction below the critical value is
considered. 

For the glass states the correlators $\phi_{3.4} (t)$ exhibit
weakly damped oscillations which lead to nearly Lorentzian
resonances for the spectra. These excitations are analogous to
phonons in crystals. The softening of the glass with 
decreasing $\varphi$
leads to a decrease of the phonon frequency, but 
Fig. \ref{fig_lin_ISF_w}
demonstrates that for all $\varphi \geq \varphi_c$ the peak
positions are located considerably above the maximum position
$\omega_P$ of the AOP discussed in the preceding paragraph. One
recognizes for the $\varphi = 0.60$ result in 
Fig. \ref{fig_lin_ISF_t} that the
oscillations for $t \leq 0.1$ do not occur around the long time
limit $f_{3.4}$. Rather the oscillation centre follows a curve
discussed above for the bouncing in the cage. This is equivalent
to the fact that the phonon resonance does not exhaust the
spectrum $\phi_{3.4}^{\biss} (\omega)$, rather it is placed on top
of some background. The background exhibits a similar
threshold as discussed above for the AOP of $\phi_{7.0}^{\biss}
(\omega)$, and with decreasing $\varphi$ it also follows the same
pattern as described for the $q = 7.0$ spectra. Apparently, the
dynamics for $q = 3.4$ illustrates a hybridization of the phonon
dynamics with the modes building the AOP. The regular
dependence of the MCT solutions on control--parameter variations
are the reason why the results in Figs. \ref{fig_lin_ISF_t} 
and \ref{fig_lin_ISF_w} do not exhibit any
drastic change if $\varphi$ is shifted from the glass state with
label $n = 4$, referring to $\epsilon = (\varphi - \varphi_c) /
\varphi_c = 0.0464$, through the critical point to the liquid
state with $\epsilon = - 0.0464$. But upon shifting the state into
the liquid the phonon resonances get buried under
the relaxation spectra.

The variation of the spectra with changes
of the wave vector $q$ is demonstrated in Fig. \ref{fig_happrox} 
for the
stiff--glass state with the packing fraction $\varphi = 0.60$. For
$q \leq 0.6$ a single peak of nearly Lorentzian shape exhausts the
whole inelastic spectrum $\phi_q^{\biss} (\omega)$. The resonance
position follows the dispersion law of high--frequency sound
$\Omega_q^{\max} = v_\infty q$. Here $v_\infty$ is the sound speed
expected from the glass susceptibility $\hat{\chi}_{q = 0}$, Eq.
(\ref{Gl4d}), $v_\infty = v_0 / \sqrt{1 - f_{q=0}} = 75.8$. 
Also the half
width of the resonance exhibits the quadratic wave--vector
variation expected for sound in an elastic continuum: $\Gamma_q =
\gamma \cdot q^2$. The single--peak shape of the spectra is found
for all wave vectors up to about $q_D /2$, where $q_D = (36 \pi
\varphi)^{1/3} = 4.08$ denotes the Debye wave vector of the
system. Also the linear dispersion law continues up to these large
$q$--values as is demonstrated in Fig. \ref{fig_Omega}. However, 
for $q \geq
0.6$ the resonance width $\Gamma_q$ is somewhat larger than
expected by extrapolating the asymptotic law $\Gamma_q = \gamma
(\Omega_q^{\max} / v_\infty)^2$, as is shown in Fig. \ref{fig_Gamma}. 
The sound
frequency $\Omega_q^{\max}$ reaches the position of the maximum of
AOP at $q \approx 1.2$ and extends considerably beyond this value
for larger $q$, as is shown for $q = 1.8$ in Fig. \ref{fig_happrox}.

If $\Omega_q^{\max}$ increases to the centre of the AOP the
Ioffe--Regel limit is approached; i.e. the sound frequency becomes
of the same order as the resonance width. In this case
hybridization of high--frequency sound with the modes forming the
AOP becomes important. The sound resonance is not Lorentzian
anymore, as demonstrated in Fig. \ref{fig_happrox}. For $q = 1.4$ 
the threshold of
the AOP near $\omega = 40$ causes a shoulder on the low--frequency
wing of the sound peak. With increasing $q$ a broad flat
background spectrum is formed between the threshold and
$\Omega_q^{\max}$ as shown in Fig. \ref{fig_happrox} for $q = 1.8$ 
and 2.2. For
wave vectors exceeding $q_D / 2$ the spectrum exhibits two maxima.
The one at lower frequency is due to the AOP. The peak at higher
frequencies is the continuation of the high--frequency sound
resonance, which is increased because of a level repulsion effect,
as demonstrated in Fig. \ref{fig_happrox} for $q$ between 2.2 and 5.4. 
The high--frequency--phonon frequency $\Omega_q^{\max}$ increases with
$q$ increasing up to about $q_D$, and then it decreases again if
$q$ increases up to about $1.5 q_D$. So the dispersion follows the
pattern expected for a crystal phonon near the Brillouin zone
boundary. The width $\Gamma_q$ of the high--frequency phonon for
$2.5 \leq q \leq 5.0$ varies only weakly with changes of $q$. Also
the AOP maximum $\omega_P$ does not change very much if $q$
increases from 2 to 5, as is demonstrated by the open circles in
Fig. \ref{fig_Omega}. In a large wave--vector interval around the 
position of
the first sharp diffraction peak of the structure factor $S_q$,
the spectra do not exhibit a phonon peak but the AOP only. The
spectral shape varies only weakly with changes of $q$, except near
$q = 8.7$; there a new splitting in two peaks occurs.

In Fig. \ref{fig_fast_sound} the spectroscopic parameters 
$\Omega_q^{\max}$ and
$\Gamma_q$ of high--frequency sound are presented as function of
the packing fraction. The resonance frequency decreases with
decreasing $\varphi$, reflecting the softening of the glass state
upon expansion. Remarkably, the damping $\Gamma_q$ does not vary
much with changes of $\varphi$.

\subsection{\bf{A generalized--hydrodynamics description for the 
glass--state dynamics}}

The description of hydrodynamic sound is obtained 
by coarse graining correlation functions so
that fluctuations on microscopic scales for the space--and--time
variations are averaged out. Coarse graining is equivalent to
restricting functions like $\hat \Omega_q$ or $\hat m_q (\omega)$ to
their leading order Taylor coefficient with respect to their
$q$--and--$\omega$ dependence. In this manner one gets from Eq.
(\ref{Gl4d}) the hydrodynamics description:
\begin{mathletters}
\label{Gl6}
\begin{equation}
\label{Gl6a}
\chi_q^H (\omega) / \hat \chi_q = \Omega_q^{H2} / \left[\omega^2 -
\Omega_q^{H2} + i \omega \Gamma_q^H \right] \,\, .
\end{equation}
Here the dispersion law $\Omega_q^H$ and the damping $\Gamma_q^H$
are given by
\begin{equation}
\label{Gl6b}
\Omega_q^H = v_\infty q \quad , \quad v_\infty^2 = v_0^2 (1 + C)
\quad , \Gamma_q^H = \gamma q^2 \quad , \quad \gamma = v_\infty^2
K^{\biss} (\omega = 0) \,\, ,
\end{equation}
\end{mathletters}
where the notation $C = C_{q = 0}$ and $K (t) = \hat m_{q = 0}
(t)$ is introduced. Because of Eqs. (\ref{Gl2a}), (\ref{Gl3}) 
and (\ref{Gl4c}) one can
express $C$ and $K(t)$ in terms of
the mode--coupling functional ${\cal{F}}_0 = {\cal{F}}_{q=0}$:
\begin{equation}
\label{Gl7}
C = {\cal{F}}_0 [f] \,\, ,\qquad K (t) = \left\{ {\cal{F}}_0 [\phi
(t)] - C \right\} / (1 + C) \,\, .
\end{equation}
An explicit expression for the new functional ${\cal{F}}_0$
follows from Eqs. (\ref{Gl2a}) and (\ref{Gl2b}):
\begin{mathletters}
\label{Gl8}
\begin{equation}
\label{Gl8a}
{\cal{F}}_0 [\tilde f] = \int_0^\infty V_k \tilde f_k^2 d k \,\, ,
\end{equation}
where the weight factors $V_k \geq 0$ read
\begin{equation}
\label{Gl8b}
V_k = \rho S_{q = 0} \left[S_k k / 2 \pi \right]^2 \left[c_k^2 +
\frac{2}{3} (k c_k^\prime) c_k + \frac{1}{5} (k c_k^\prime)^2
\right] \,\, .
\end{equation}
\end{mathletters}
The integral in Eq. (\ref{Gl8a}) is to be understood as a Riemann sum over
the wave--vector grid of $M$ terms as specified in Sec.IIA.

The hydrodynamics description is not suited to deal with
high--frequency sound and the AOP. In order to identify the
essence of these phenomena a generalized--hydrodynamics
description shall be developed. It is obtained from Eq. (\ref{Gl4d}) via
approximating kernel $\hat m_q (\omega)$ by its zero--wave--vector
limit:
\begin{equation}
\label{Gl9}
\chi_q^{GHD} (\omega) / \hat\chi_q =  \hat\Omega_q^2 / \left[
\omega^2 - \hat\Omega_q^2 (1-\omega K (\omega))\right] \,\, .
\end{equation}
This formula extends the hydrodynamics approximation in two
respects. Firstly and most importantly, the hydrodynamic damping
constant $i \gamma / v_\infty^2$ is replaced by the
frequency--dependent function $K (\omega) = K^\prime (\omega) + i
K^{\biss} (\omega)$. Secondly, the full non--trivial
$q$ dependence of $\hat \Omega_q$ is kept. According to 
Fig. \ref{fig_Omega} it
is the $q$--dependence of  $\hat \Omega_q$ which dominates that of
the whole density--fluctuation spectra. If one intends to describe
these spectra for wave vectors up to the inverse of the
interparticle distances, one must not simplify $\hat \Omega_q$.
However, if one is merely interested in the description of
high--frequency sound so that $q$ is restricted to the regime
below $q_D /2$, say, one can replace $\hat \Omega_q$ in Eq. (\ref{Gl9}) by
the hydrodynamics limit $\Omega_q^H = v_\infty q$. In this case,
formula (\ref{Gl9}) has a form often used in acoustics, where $(1 - \omega
K (\omega))$ is called the dimensionless longitudinal elastic
modulus. Let us also rewrite the mode--coupling formula (\ref{Gl7}) for
$K(t)$ more transparently in analogy to Eqs. (\ref{Gl5}):
\begin{mathletters}
\label{Gl10}
\begin{equation}
\label{Gl10a}
K (t) = K^{(1)} (t) + K^{(2)} (t) \,\, ,
\end{equation}
\begin{equation}
\label{Gl10b}
K^{(1)} (t) = \int_0^\infty W_k^{(1)} \hat\phi_k (t) \quad , \quad
W_k^{(1)} = 2 (1 - f_{q=0}) V_k f_k (1 - f_k) \,\, ,
\end{equation}
\begin{equation}
\label{Gl10c}
K^{(2)} (t) = \int_0^\infty W_k^{(2)} \hat\phi_k (t)^2 \quad ,
\quad W_k^{(2)} = (1 - f_{q=0}) V_k (1 - f_k)^2 \,\, .
\end{equation}
\end{mathletters}

In Fig. \ref{fig_happrox} representative MCT spectra for the 
glass states of the
HSS are compared with the results of the
generalized--hydrodynamics approximation. For the regime $q \leq
2$ the description by Eq. (\ref{Gl9}) with $\hat \Omega_q = v_\infty q$ is
very good. This does not only hold for the treatment of the
high--frequency--sound resonance but also for the background
spectrum. But the generalized--hydrodynamics approximation yields
reasonable results also for wave vectors extending up to and
exceeding the value of the structure--factor--peak position
$q_{\max} \approx 7$. In particular the subtle hybridization of
the phonon with the modes underlying the AOP are treated
semi quantitatively correctly. 
Hence, one concludes that Eqs. (\ref{Gl9}) and
(\ref{Gl10}) separate the problem of the evolution of the transient
dynamics into two parts. The first part concerns the dependence on
the wave vectors. This is determined entirely by the quantities
$\hat \chi_q$ and $\hat \Omega_q$. These functions are constructed
from the structure factor $S_q$, whose dependence on $q$ and
$\varphi$ is well understood. $S_q$ determines the
characteristic frequency $\Omega_q$ and the thermodynamic
susceptibility $\chi_q^T$. The MCT equations provide the glass
form factors $f_q$, which are understood as well
\cite{Bengtzelius84}. And the combination of this information
yields $\hat \Omega_q$ and $\hat \chi_q$ via Eqs. (\ref{Gl4b}) 
and (\ref{Gl4d}).
Consequently, the remaining issue of this paper is to provide the
understanding of the kernel $K(\omega)$.

Figure \ref{fig_modulus} exhibits the spectra $K^{\biss} (\omega)$ 
and also the
reactive parts of the modulus $\Delta (\omega) = 1 - \omega
K^\prime (\omega)$ for three representative glass states. 
The spectra $K^{\biss} (\omega)$ are much simpler than those for
the correlators since phonon resonances are absent. This property
is the essential result achieved by the Zwanzig--Mori theory, and
MCT preserves this property. The spectra of the kernels consist of
some broad background and an AOP. The AOP of $K^{\biss} (\omega)$
has quite a similar form as discussed in Sec.IIIA for
$\phi_{7.0}^{\biss} (\omega)$ and also its changes due to changes
of $\varphi$ are similar. Formula (\ref{Gl9}) describes the hybridization
of two "oscillators". One is a bare phonon with dispersion $\hat
\Omega_q$ and the other is represented by the AOP spectrum. The
hybridization problem is analogous to the one considered, for
example, in the dynamical theory of light propagation in
dielectric media. The bare phonon corresponds to the
electro--magnetic wave in vacuum and $1 /[1 - \omega K (\omega)]$
is the analogue of the dielectric function. In lowest order one
gets two resonances of the coupled system whose frequencies
$\Omega_q^{\max}$ are obtained approximately as solution of the
equation $(\Omega_q^{\max} / \hat \Omega_q)^2 = \Delta
(\hat\Omega_q^{\max})$ and the damping is estimated by $\Gamma_q =
\hat\Omega_q^2 K^{\biss} (\hat \Omega_q^{\max})$. Elementary
discussion of these equations with the aid of Fig. \ref{fig_modulus}, 
which is
left to the reader, explains qualitatively the findings reported
in Figs. \ref{fig_happrox}-\ref{fig_fast_sound}.

A digression might be useful concerning the different role played
by the two equations for the susceptibility (\ref{Gl1b}) and (\ref{Gl4d}). 
The crucial property of glassy dynamics is that kernel $m_q (\omega)$
in Eq.(\ref{Gl1b}) depends strongly on frequency $\omega$. In particular,
this kernel has a large reactive part $m_q^\prime (\omega)$.
Therefore it makes no sense to try a hydrodynamics approximation
for Eq. (\ref{Gl1b}), based on $m_q (\omega) \approx m_q (\omega = 0) = i
m_q^{\biss} (\omega = 0)$. But let us consider a
generalized--hydrodynamics approximation, defined by ignoring the
wave vector dependence of kernel $m_q (\omega)$. Because of Eq.
(\ref{Gl2a}) this limit is given by ${\cal{F}}_0 [\phi (t)]$ and thus it
can be expressed by $K (\omega)$ via Eq. (\ref{Gl7}):
\begin{equation}
\label{Gl11}
m_{q=0} (\omega) = \left[ - C /\omega \right] + ( 1 + C) K(\omega)
\,\, .
\end{equation}
The kernel $m_{q=0} (\omega)$ exhibits the pole $[- C / \omega]$,
which is the signature of the ideal glass state; other subtleties
of the glassy dynamics are hidden in $K (\omega)$. Substitution of
Eq. (\ref{Gl11}) for $m_q (\omega)$ in Eq. (\ref{Gl1b}) reproduces 
Eq. (\ref{Gl9}), except
that the renormalized frequency $\hat\Omega_q$ is replaced by the
approximation to Eq. (\ref{Gl4b}): $\Omega_q / \sqrt{1 - f_{q=0}}$. For
our model of the HSS structure there are no serious wave--vector
dependencies of $f_q$ for $q \leq q_D / 2$. This means that
high--frequency sound and the background spectrum for $q \leq 2$
can also be discussed on the basis of Eq. (\ref{Gl1b}). The dotted lines
in Fig. \ref{fig_happrox} exhibit the corresponding results for 
$q = 1.0$ and 3.4.
However, Fig. \ref{fig_DWF} demonstrates that the relevant factor 
$1 - f_q = 1/ (1 + C_q)$ varies by more than 100\% for $q$ approaching 
and exceeding $q_D$. Therefore the generalized--hydrodynamics
approximation based on Eq. (\ref{Gl11}) cannot be used to discuss, for
example, $\phi_{7.0}^{\biss} (\omega)$. The superiority of the
generalized--hydrodynamics approximation based on Eq. (\ref{Gl4d}) rather
than on Eqs. (\ref{Gl1b}) and (\ref{Gl11}) results from the fact, that 
Eqs. (\ref{Gl9})
and (\ref{Gl10}) treat the glass structure as it comes out from 
Eq. (\ref{Gl3}),
thus avoiding the additional small--$q$ approximation $C_q \approx
C_{q=0}$.

\subsection{\bf{The stiff--glass approximation}}

In this section Eqs. (\ref{Gl1}) and (\ref{Gl2}) shall be
considered for large coupling coefficients $V_{q,kp}$. Let us
write symbolically $V_{q,kp} = O(1/\eta)$, so that various
quantities can be classified according to their power of the small
parameter $\eta$. For the function $C_q$, which enters Eq. (\ref{Gl3}) for 
$f_q$, one gets large values: $C_q =
O(1/\eta)$. Thus $1 - f_q = O(\eta)$. Stiff glass states are
characterized by Debye--Waller factors close to unity, as is
demonstrated by curve (a) in Fig. \ref{fig_DWF}. Therefore one concludes 
from Eqs. (\ref{Gl5b}), (\ref{Gl5c}), (\ref{Gl10b}) and (\ref{Gl10c}) 
that the renormalized coupling
coefficients decrease towards zero in the limit $\eta \to 0$:
\begin{equation}
\label{Gl12}
\hat V_{qk} , W_k^{(1)} = O(\eta) \, , \quad \hat V_{q,kp} ,
W_k^{(2)} = O(\eta^2) \,\, .
\end{equation}
After eliminating the arrested glass structure as 
described in Sec.IIB, the remaining MCT equations (\ref{Gl4d})
and (\ref{Gl5}) deal with a weak--coupling situation if the system is in a
deep glass state. Figure \ref{fig_V0} demonstrates how the coefficients
$W_q^{(1,2)}$ decrease with increasing packing fraction. Let us
introduce for later use also the integrated coupling coefficients
\begin{equation}
\label{Gl13}
w_{1,2} = \int_0^\infty W_k^{(1,2)} dk \,\, .
\end{equation}
For the three packing fractions $\varphi$ = 0.600 (0.567, 0.540)
dealt with in Fig. \ref{fig_V0} one finds $w_1$ = 0.39 (0.53, 0.64) and
$w_2$ = 0.14 (0.34, 0.78) respectively. Let us recall from Sec.IIB
that $\hat V_{q,k}$ and the derived quantities $W_{k}^{(1)}, w_1$
specify the interactions of density fluctuations with the arrested
amorphous structure, while $\hat V_{q,kp},
W_k^{(2)}, w_2$ quantify two--mode--decay processes. One concludes
furthermore that in the limit $\eta \to 0$ the two--mode
contributions to the kernel $m_q (\omega)$ get suppressed relative
to the one--mode contributions, in particular: $K^{(1)} (t) = O
(\eta), \, K^{(2)} (t) = O (\eta^2)$. This explains the result
shown in Fig. \ref{fig_scat_cont}: for the packing fraction 
$\varphi = 0.60$ the
one--mode contribution $K^{(1)\biss} (\omega)$ provides the
dominant part of the total spectrum $K^{\biss} (\omega)$. The 
role played by the two
contributions $K^{(1)}$ and $K^{(2)}$ is utterly different. The
former yields the AOP and the latter provides a background spectrum.

To explain the AOP a stiff--glass approximation shall be analysed
which is indicated by superscripts (1). It is defined by dropping
the two--mode contribution to $\hat m_q (\omega)$ in Eq. (\ref{Gl4d}).
Substituting the result into the formula $\phi_q^{(1)} (\omega) =
[\chi_q^{(1)} (\omega) / \hat \chi_q - 1] /\omega$ one gets
\begin{mathletters}
\label{Gl14}
\begin{equation}
\label{Gl14a}
{\phi}_q^{(1)} (\omega) = - 1 / \left[ \omega - \hat\Omega_q^2 /
\left[ \omega + i\nu_q + \hat\Omega_q^2 m_q^{(1)} (\omega) \right]
\right] \,\, ,
\end{equation}
\begin{equation}
\label{Gl14b}
m_q^{(1)} (\omega) = \hat{\cal{F}}_q^{(1)} \left[ \phi^{(1)}
(\omega)\right] \,\, .
\end{equation}
\end{mathletters}
In order to ease the discussion at the end of this section, a
friction term proportional to $\nu_q \geq 0$ has been included in
the formulas. It is equivalent to complementing $m_q^{(1) \biss}
(\omega)$ by a white noise spectrum. Unless emphasized otherwise,
one may read the formulas with $\nu_q = 0$. Equations (\ref{Gl14}) 
define
a special model for the MCT and all general theorems quoted in
Sec.IIA apply. If one would treat the mode--coupling coefficients
$\hat V_{qk} \geq 0$ in the Eq. (\ref{Gl5b}) for the functional
$\hat{\cal{F}}_q^{(1)}$ as free parameters, Eqs. (\ref{Gl14}) could
exhibit glass transitions. For the discussion of such bifurcations
the $M$ by $M$ matrix $C_{qk} = \left\{\partial{\cal{F}}_q [f] /
\partial f_k \right\} (1 - f_k)^2, \, q,k = 1,2, \ldots M$, which is
called the stability matrix, plays an essential role. Glass
transitions are characterized by the spectral radius E of matrix C
to be unity, and for all other states one gets $E < 1$
\cite{Goetze91b,Goetze95b}. One checks, that the stability matrix
of the complete theory as defined by Eqs. (\ref{Gl4}) and (\ref{Gl5}) 
is the same
as the one for the stiff--glass approximation, defined by Eqs.
(\ref{Gl14}). Therefore Eqs. (\ref{Gl14}) with the coefficients 
$\hat V_{qk}$
defined in Eq. (\ref{Gl5b}) do not exhibit glass--transition points
anymore; in particular $\phi_q^{(1)} (t \to \infty) = 0$. Let us
note also that matrix C is equivalent to matrix $\hat V : C_{qk} =
(1 - f_q) \hat V_{qk} / ( 1 - f_k)$. Therefore also the spectral
radius of $\hat V$ is $E$. Hence the resolvent $S = (1 - \hat
V)^{-1}$ exists.

The $\omega = 0$ limit of Eqs. (\ref{Gl14}) yields the linear equation
$\sum_k (\delta_{qk} - \hat V_{qk}) m_k^{(1)} (\omega = 0) = i
\sum_k \hat V_{qk} \nu_k / \hat \Omega_k^2$. For the relevant case
$\nu_k = 0$ one concludes that the zero--frequency spectrum of the
kernel vanishes. Therefore 
\begin{mathletters}
\label{Gl15}
\begin{equation}
\label{Gl15a}
m_q^{(1) \biss} (\omega \to 0) = R_q \omega^2  \,\, ,
\end{equation}
and this implies for the low frequency behavior of the correlator
\begin{equation}
\label{Gl15b}
\phi_q^{(1) \biss} (\omega \to 0) = (\pi/2) \left[\delta (\omega -
\Omega_q^H) + \delta (\omega + \Omega_q^H) \right] + R_q \omega^2
\,\, .
\end{equation}
\end{mathletters}
Hence, one finds Rayleigh's law: the scattering probability
$m_q^{(1) \biss} (\omega)$ of low--frequency phonons from static
density fluctuations varies proportional to $\omega^2$.

Let us simplify the stiff--glass approximation by embedding it
into the generalized--hydrodynamics description. According to
Sec.IIIB this amounts to approximating $m_q^{(1)} (\omega)$ by its
$q = 0$ value, denoted $K^{(1)} (\omega)$. Similarly
one should write for the so far not specified friction coefficient
$\nu_q = \tau \hat\Omega_q^2$. Let us denote the stiff--glass
susceptibility resulting from Eq. (\ref{Gl14a}) within the
generalized--hydrodynamics approximation by $\chi_q^{SGA} (\omega)
= \hat\chi_q (\phi_q^{(1)} (\omega) \cdot \omega + 1)$, so that
\begin{equation}
\label{Gl16}
\chi_q^{SGA} (\omega) = \hat\chi_q \hat\Omega_q^2 / \left[\omega^2
- \hat\Omega_q^2 + \omega \hat\Omega_q^2 \left[ i \tau + K^{SGA}
(\omega)\right] \right] \,\, .
\end{equation}
The problem is reduced to evaluating the kernel $K^{SGA} (\omega)$
from the equation
\begin{equation}
\label{Gl17}
K^{SGA} (\omega) = - \int_0^\infty dk W_k^{(1)} \{ 1 /
\left[\omega - \hat\Omega_k^2 / \left[\omega + \hat\Omega_k^2 (i
\tau + K^{SGA} (\omega)\right] \right]\} \,\, .
\end{equation}
Again, the general theorems of MCT apply. However, since the mode
coupling coefficients have been altered by substituting $\hat
V_{0k}$ for $\hat V_{qk}$, the stability matrix has been changed.
Therefore it cannot be excluded that the spectral radius E reaches
unity for the simplified theory . This would be an artifact of the
generalized--hydrodynamics approximation and Eqs. (\ref{Gl16}) 
and (\ref{Gl17})
must not be used in such case.

The generalized--hydrodynamics approximation simplifies the Eq.
(\ref{Gl15a}) by the replacement of $R_q$ by its $q = 0$ limit to be
denoted by $R_0 : K^{SGA \biss} (\omega \to 0) = R_0 \omega^2$.
The coefficient $R_0$ is obtained by substituting Eq.
(\ref{Gl15b}) into Eq. (\ref{Gl10b}). Using Eq. (\ref{Gl8b}) 
one gets
\begin{equation}
\label{Gl18}
R_0 = \rho (S_0 / v_\infty)^3 c_0^2 f_0 (1 - f_0)^2 / \left[ 4
\pi^2 (1 - w_1)\right] \,\, .
\end{equation}
The Rayleigh spectrum is included in Fig. \ref{fig_scat_cont}. 
A huge magnification
was necessary to make this contribution visible on the scale of
$K^{\biss} (\omega)$.

One infers from Fig. \ref{fig_Omega} that for $q > 2$ the variations 
of the renormalized frequency $\hat\Omega_q$ with changes of the wave
vector $q$ is suppressed relative to the ones of $\Omega_q$. One
can get an estimation of the $k > q_D$ contribution to the
integral (\ref{Gl17}) by replacing $\hat \Omega_k$ by some averaged value
$\tilde\Omega$. Figure \ref{fig_Omega} suggests $\tilde \Omega = 120$ for
$\varphi = 0.60$. Introducing $\nu = \tau \tilde\Omega^2$ and
indicating the results of the specified estimation by a tilde, Eq.
(\ref{Gl16}) is equivalent to 
\begin{mathletters}
\label{Gl19}
\begin{equation}
\label{Gl19a}
\phi_{q \geq q_D}^{SGA} (\omega) \approx \tilde\phi (\omega) = - 1
/ \left[\omega - \tilde\Omega^2 / \left[\omega + i \nu +
\tilde\Omega^2 \tilde K (\omega)\right] \right] \,\, .
\end{equation}
According to Fig. \ref{fig_V0} the $k \leq q_D$ contributions to the
integral in Eq. (\ref{Gl17}) can be neglected, and thus this equation for
the kernel simplifies to
\begin{equation}
\label{Gl19b}
K^{SGA} (\omega) \approx \tilde K (\omega) = w_1  \tilde \phi
(\omega)   \,\, .
\end{equation}
\end{mathletters}
The solution of Eqs. (\ref{Gl19}) for $\tilde K$ and $\tilde \phi$ reads
\begin{mathletters}
\label{Gl20}
\begin{equation}
\label{Gl20a}
\tilde K (\omega) = \left[\omega_+ \omega_- - z (\omega)^2 +
\sqrt{z (\omega)^2 - \omega_-^2} \cdot \sqrt{z (\omega)^2 -
\omega_+^2} \right] / (2 \omega) \,\, ,
\end{equation}
\begin{equation}
\label{Gl20b}
z (\omega)^2 = \omega (\omega + i \nu) / \tilde\Omega^2  \,\, ,
\qquad \omega_{\pm} = 1 \pm \sqrt{w_1} \,\, .
\end{equation}
\end{mathletters}
The coupling constant $w_1$ takes over the role of the spectral
radius of the stability matrix. In order avoid artifacts of the
various approximations leading to the results (\ref{Gl20}), the formulas
can be applied only for $w_1 \leq 1$. Figure \ref{fig_mz} demonstrates that
Eqs. (\ref{Gl20}) with $\nu = 0$ describe the AOP of $K^{\biss}(\omega)$
for the HSS with $\varphi = 0.60$ reasonably well. One could
improve the description by trying better choices for $\tilde
\Omega$, but this would not lead to any new insight.

Figure \ref{fig_V0} shows that also for the evaluation of the integral in
Eq. (\ref{Gl10c}) the contributions for $k < q_D$ can be ignored. A
leading estimation for the two--mode kernel can thus be obtained
as $\tilde K^{(2)} (t) = w_2 \tilde \phi (t)^2$. Using Eq. (\ref{Gl19b})
one gets therefore
\begin{equation}
\label{Gl21}
\tilde K^{(2)} (\omega) = (1 / \pi) (w_2 / w_1^2) \int \tilde K
(\omega - \omega^\prime) \tilde K^{\biss} (\omega^\prime) d
\omega^\prime \,\, .
\end{equation}
The zero--frequency limit leads to a trivial integral with the
result $\tilde K^{(2) \biss} (\omega = 0) = \tau = (w_2 /
\sqrt{w_1}) 8 /(3 \pi \tilde \Omega) $. Going back to Eq. (\ref{Gl10a})
one concludes: the background term due to $K^{(2)} (t)$ could have
been taken into account in its white noise approximation. This
leads to the extension of the equation of motion by adding a
friction term $\nu = \tilde \Omega^2 \cdot \tau$. Such an
extension is obtained by including this term in the formulas, as
was done already in Eq. (\ref{Gl14a}) and the following formulas. The
dotted line in Fig. \ref{fig_mz} demonstrates that thereby all qualitative
features of $K^{\biss}(\omega)$ are understood. 
A further improvement is obtained by
dropping the white noise approximation for the correction term
$\tilde K^{(2)}$. This is done by replacing $\omega + i \nu$ by
$\omega + \tilde K^{(2)} (\omega)$ in Eq. (\ref{Gl20b}). Figure \ref{fig_mz}
demonstrates that thereby a more satisfactory treatment of
$K^{\biss} (\omega)$ is obtained.

The periodic continued fraction for $\tilde \phi (\omega)$, which
is defined by Eqs. (\ref{Gl19}), can be related to a Hilbert--Stieltjes
integral: $\int_{-1}^1 dx \sqrt{1 - x^2} / (x - \zeta) = \pi (-
\zeta + \sqrt{\zeta - 1} \sqrt{\zeta + 1})$. Thus one can express
the normalized susceptibility $\tilde \chi (\omega) = [\omega
\tilde \phi (\omega) + 1]$ in the form:
\begin{mathletters}
\label{Gl22}
\begin{equation}
\label{Gl22a}
\tilde \chi (\omega) = \int_{\omega_-^2}^{\omega_+^2} d \xi \tilde
\rho (\xi) \chi_\xi (\omega) \,\, ,
\end{equation}
\begin{equation}
\label{Gl22b}
\tilde \rho (\xi) = \sqrt{(\omega_+^2 - \xi)(\xi - \omega_-^2)} /
(2 \pi w_1) \,\, ,
\end{equation}
\begin{equation}
\label{Gl22c}
\chi_\xi (\omega) = - \tilde \Omega^2 / \left[ \omega^2 - \tilde
\Omega^2 \xi + i \omega \nu \right] \,\, .
\end{equation}
\end{mathletters}
For the stiff--glass states the AOP is obtained as a superposition
of undamped--harmonic--oscillator spectra. The weight distribution
$\tilde\rho (\xi)$ for the oscillators with frequency $\sqrt{\xi}
\tilde \Omega$ extends from $\Omega_- = \omega_- \tilde\Omega$ to
$\Omega_+ = \omega_+ \tilde \Omega$. If the approximate
description is extended so that two--mode interactions are
incorporated as white--noise--background spectrum for the
fluctuating--force kernels, the results remain valid, but the
oscillator dynamics has to include a Newtonian friction term,
quantified by $\nu \geq 0$. A better description of the spectra
for frequencies large compared to $\Omega_-$ is obtained by
acknowledging that the friction forces do not exhibit a white
noise spectrum. This can be done by replacing $i \nu$ in Eq. (\ref{Gl22c})
by the kernel $\tilde K^{(2)} (\omega)$ from Eq. (\ref{Gl21}). The glass
instability for $T \to T_c$ or $\varphi \to \varphi_c$ is
connected with the approach of $w_1$ to unity, i.e., with the
threshold $\Omega_-$ approaching zero.

\section{Schematic--Model Discussions}
\subsection{\bf{Models for anomalous--oscillation peaks}}

The simplest MCT models deal with a single correlator, say
$\phi (t)$. The Eq. (\ref{Gl1a}) remains valid with the
subscript $q$ dropped. Generalizing Eq. (\ref{Gl2a}), 
the kernel
$m (t)$ is written as mode--coupling function, specified by a
series with coefficients $v_n \geq 0$:
\begin{equation}
\label{Gl23}
m (t) = {\cal{F}} \left[ \phi (t) \right] = \sum_{n = 1}^\infty
v_n \phi (t)^n \,\, .
\end{equation}
The long--time limit $f = \phi (t \to \infty)$ obeys Eq. (\ref{Gl3}): $f =
{\cal{F}} [f] / (1 + {\cal{F}} [f])$. For the glass states one can
carry out the transformation discussed in Sec.IIB to get
correlators $\hat \phi (t)$ and kernels $\hat m (t)$ with
vanishing long time limits. Equations (\ref{Gl4}) and (\ref{Gl5a}) 
remain valid
with the subscript $q$ dropped. The new functional $\hat
{\cal{F}}$ is again a power series. For the renormalized Taylor
coefficients one gets in analogy to Eqs. (\ref{Gl5b}) and (\ref{Gl5c}): 
$\hat v_n
= (1-f)^{n+1} [\partial^n {\cal{F}} [f] /
\partial f^n] / n ! , \, n = 1, 2, \ldots$. The limit of a stiff
glass is obtained if at least one of the coefficients $v_n$
becomes large. The number $\eta = 1 - f$ can be used as small
parameter for the classification of terms. 
One finds $\hat v_n = O (\eta^n)$. Let
us change the notation to $w_{1,2} = \hat v_{1,2}$ so that
\begin{equation}
\label{Gl24}
w_1 = (1 - f)^2 \sum_{n = 1}^\infty n v_n f^{n - 1} \,\, , \qquad
w_2 = (1 - f)^3 \sum_{n = 2}^\infty n (n - 1) v_n f^{n - 2} \,\, .
\end{equation}
The stiff--glass approximation can be defined as in Sec.IIIC by
approximating the kernel by the leading term $\hat m (t) = w_1
\hat\phi (t)$. Let us change the notation to $\hat\phi =
\tilde\phi$ and $\hat\Omega = \tilde\Omega$, so that
\begin{equation}
\label{Gl25}
\phi (t) = f + (1 - f) \tilde\phi (t) \,\, , \qquad \tilde\Omega^2
= \Omega^2 / (1-f) \,\, .
\end{equation}
With $\tilde K (t) = \hat m (t)$ one arrives at the pair of
equations (\ref{Gl19}).

Let us consider as an example the model specified by the
functional ${\cal{F}} [f] = v_1 f + v_2 f^2$. It was introduced
as the simplest one which can reproduce all possible
anomalous exponents of the general
MCT \cite{Goetze91b,Goetze84}. Liquid--glass transitions occur on
a line in the $v_1 - v_2$ parameter plane. The line, where the
long--time limit $f$ jumps from zero to $f^c = 1 - \lambda > 0$,
is a piece of a parabola with the representation $v_1^c = (2
\lambda - 1) / \lambda^2, \, v_2^c = 1 /\lambda^2 , \, 0.5 \leq
\lambda < 1$. In Ref. \cite{Goetze96b} diagrams analogues to Figs.
\ref{fig_HSM_ISF_t} and \ref{fig_HSM_ISF_w} can be found, 
which exhibit the evolution of glassy
dynamics and the AOP upon shifting $(v_1, v_2)$ from the
weak--coupling regime to the strong--coupling one. As path in the
parameter plane a straight line was chosen: $v_{1,2} = v_{1,2}^c
(1 + \epsilon), \, \lambda = 0.7, \, \epsilon = \pm 1 / 4^n, \, n
= 0, 1, \ldots$ The full lines in Fig. \ref{fig_F12_deep} reproduce three
glass--state results. They are similar to what is
demonstrated for the HSS in the lower panel of Fig. \ref{fig_lin_ISF_w}. 
The reason
was explained in Sec.IIIC: these HSS spectra are described by the
stiff--glass approximation and this approach yields the same
formulas (\ref{Gl19}) as derived above for the schematic model. 
The dashed
lines in Fig. \ref{fig_F12_deep} represent the leading order 
description by Eqs.
(\ref{Gl19}) with $\nu = 0$ and $w_1, \tilde\Omega$ determined by 
Eqs. (\ref{Gl24}) and (\ref{Gl25}). The dotted lines incorporate 
$\nu = \tilde \Omega^2
\tilde K^{(2) \biss} (\omega = 0)$ evaluated from Eq. (\ref{Gl21}). 
The results for the $n = 4$ spectra show that the extended
stiff--glass approximation can describe the dynamics reasonably
well for states which differ from the instability point by only
6.25\%. It was pointed out \cite{Goetze96b} that the
schematic--model solutions in Fig. \ref{fig_F12_deep} fit 
qualitatively to the
boson--peak scenario as reported for neutron--scattering data. It
was shown \cite{Das99} that the AOP of the specified model
can be used to fit the boson--peak spectrum of orthoterphenyl as
it was measured \cite{Sokolov94} by Raman scattering at $T = 245K
= T_c - 45K$ for frequencies between 100 GHz and 700 GHz.

Usually, a one--component schematic model is too restrictive to
deal quantitatively with experimental data. But one can construct
more elaborate schematic models with the intention to mimic more
features of the MCT. The perspective of such approach shall be
indicated by results for the model used recently for the
interpretation of scattering spectra of glassy liquids
\cite{Franosch97a,Singh98,Ruffle97,Ruffle98,Ruffle99}. The model
extends the one of the preceding paragraph by introducing a second
correlator, to be denoted by $\phi^s (t)$. The equation of motion
has the standard form of Eq. (\ref{Gl1a}) with $\Omega_q, \phi_q$, and
$m_q$ replaced by $\Omega^s, \phi^s$, and $m^s$ respectively. The
mode--coupling functional for $m^s$ is characterized by a single
coupling constant $v_s \geq 0$:
\begin{equation}
\label{Gl26}
m^s (t) = v_s \phi (t) \phi^s (t) \,\, .
\end{equation}
The model was introduced for the description of
tagged--particle motion in liquids \cite{Sjoegren86}.

The long--time limit $f^s = \phi^s (t \to \infty)$ is obtained
from Eq. (\ref{Gl3}) as $f^s = 1 - 1 / ( f v_s)$. The transformation 
to a new correlator and a new kernel with vanishing long time limits
can be done as explained in Sec.IIB: $\tilde\Omega^{s2} =
\Omega^{s2} / (1 - f^s) ; \, \hat\phi^s (t) =[\phi^s (t) - f^s] /
(1 - f^s)$. From Eq. (\ref{Gl4c}) one gets
\begin{mathletters}
\label{Gl27}
\begin{equation}
\label{Gl27a}
\hat m^s (t) = K_s^{(1)} (t) +  K_s^{(2)} (t) \,\, ,
\end{equation}
\begin{equation}
\label{Gl27b}
K_s^{(1)} (t) = u \hat\phi (t) + w_s \hat\phi^s (t) \qquad ,
\qquad K_s^{(2)} (t) = \hat v_s \hat\phi (t) \hat\phi^s (t) \,\, ,
\end{equation}
\begin{equation}
\label{Gl27c}
u = (1 - f) f^s / f \quad , \quad w_s = (1 - f^s) \quad , \quad
\hat v_s = (1 - f^s)(1 - f) / f \,\, .
\end{equation}
\end{mathletters}

The stiff--glass approximation for the two component model
requires $(1 - f)$ and $(1 - f^s)$ to be small so that $K_s^{(2)}
(t)$ can be neglected compared to $K_s^{(1)} (t)$. Let us denote
the results by a tilde. The equations of motion, which specialize
Eqs. (\ref{Gl14}), read:
\begin{mathletters}
\label{Gl28}
\begin{equation}
\label{Gl28a}
\tilde\phi^s (\omega) = - 1 / \left[ \omega - \tilde\Omega^{s2} /
\left[ \omega + \varphi (\omega) + \tilde \Omega_s^2 \tilde K^s
(\omega) \right] \right] \,\, ,
\end{equation}
\begin{equation}
\label{Gl28b}
\tilde K^s (\omega) = w_s \tilde\phi^s (\omega) \qquad , \qquad
\varphi (\omega) = \tilde\Omega^{s2} u \tilde\phi (\omega) \,\, .
\end{equation}
\end{mathletters}
This result is equivalent to Eqs. (\ref{Gl19}) except that the friction
constant $i \nu$ is generalized to a friction function $\varphi
(\omega)$. The solution can therefore be written in form of Eq.
(\ref{Gl20a}) with $z (\omega)^2$ and $\omega_{\pm}$ replaced,
respectively, by
\begin{equation}
\label{Gl29}
z^s (\omega)^2 = \omega (\omega + \varphi (\omega)) /
\tilde\Omega^{s2} \,\, , \qquad \omega_{\pm}^s = 1 \pm \sqrt{w_s}
\,\, .
\end{equation}
In particular Eqs. (\ref{Gl22}) hold with the appropriate change in the
notation. The susceptibility is a superposition with weight
$\tilde \rho^s (\xi)$ of the functions
\begin{equation}
\label{Gl30}
\chi_\xi^s (\omega) = - \tilde \Omega^{s2} / \left[ \omega^2 -
\tilde \Omega^{s2} \xi + \omega \varphi (\omega) \right] \,\, .
\end{equation}
This is a harmonic--oscillator susceptibility where the
interaction with the background is included via a friction
function $\varphi (\omega)$, dealing with the AOP of the
surrounding.

Figure \ref{fig_FS_deep} shows solutions for three characteristic 
choices of
$\Omega^s$. The results refer to the stiff glass--state discussed
in Fig. \ref{fig_F12_deep} for the label $n = 0$. The coupling 
$v^s = 10 / f$ was
chosen so that $1 - f^s = 0.10$. The stiff--glass approximation
results are shown as dashed lines. For $\Omega^s = 0.15 \Omega$ 
the resonance of the second correlator is located so far
below the AOP of the first correlator, that there appears an AOP
of $\phi^{s \biss} (\omega)$ quite similar as discussed for the
one--component model. In this case $\varphi (\omega)$ in Eq. (\ref{Gl30})
only produces a renormalization of the frequencies $\tilde\Omega^s
\sqrt{\xi}$ and a strongly suppressed background. For $\Omega^s =
\Omega$ the resonances of $\chi_\xi^s$ are shifted upward because
of level repulsion and the hybridization yields a broad background
extending from the threshold of the AOP of the first correlator to
the AOP position of the second one. For $\Omega^s = 0.6 \Omega$
the spectrum $\phi^{s \biss} (\omega)$ exhibits an AOP whose low
frequency threshold and maximum position are close to the ones for
the AOP of $\phi^{\biss} (\omega)$. The hybridization causes a
suppression of the high frequency spectrum. Therefore the AOP in
$\phi^{s \biss} (\omega)$ is more asymmetrical than the peak in
$\phi^{\biss} (\omega)$.
The hybridization results have similarities to
the ones discussed in connection with Fig. \ref{fig_happrox}. Indeed, 
it was
explained in Sec.IIIC that the phonon modes are influenced by the
large--wave--vector modes which built the AOP, but that there is
no feedback of the phonons for wave vector $q < q_D$ on the AOP.
This is the same situation as treated by the specified schematic
two--component model.

\subsection{\bf{Random--oscillator models}}

To get more insight into Eqs. (\ref{Gl22}), the following problem shall be
considered: evaluate the averaged dynamical susceptibility $\chi
(\omega)$ of an ensemble of independent harmonic oscillators. The
oscillators are specified by their mass $\mu$ and by their
frequencies $\Omega (\xi) > 0$. The latter depend on a random
variable $\xi$. Its distribution shall be denoted by $\rho (\xi) ;
\, \rho (\xi) \geq 0, \, \int \rho (\xi) d\xi = 1$. It is no
restriction of generality to assume 
$\Omega (\xi)^2 = \alpha + \beta \xi, \, \beta > 0$; but
$\xi$ has to be restricted from below to insure stability. Let us
choose as minimum for $\xi$ the value -1. Thus one can write
$\Omega (\xi)^2 = \Omega_0^2 [1 + w + 2 \sqrt{w} \xi]$, so that
$\Omega_0 > 0$ defines the frequency scale and $w , 0 \leq w < 1$,
characterizes the minimum frequency $\Omega_- = \Omega_0 (1 -
\sqrt{w})$. Denoting averages by bars, $\overline{A (\xi)} = \int
A(\xi) \rho(\xi) d\xi$, the quantity of interest is
\begin{equation}
\label{Gl31}
\chi (\omega) = \overline{- \mu / \left[ \omega^2 - \Omega (\xi)^2
\right]} \,\, .
\end{equation}
Let us define a characteristic frequency $\Omega > 0$ by
$\Omega^{-2} = \overline{\Omega (\xi)^{-2}}$. It specifies the
static susceptibility $\hat \chi = \mu / \Omega^2$. If brackets
denote canonical averaging, defined with respect to the oscillator
Hamiltonian $H_\xi = [(P^2 / \mu) + \mu \Omega (\xi)^2 Q^2] /2$,
one gets for the fluctuations of the momentum $\langle P^2 \rangle
= \mu^2 v^2$ and of the displacement $\langle Q^2 \rangle = v^2 /
\Omega (\xi)^2$, where $v$ denotes the thermal velocity. The time
evolution of some variable $A = A(Q,P)$ can be written as usual in
terms of a Liouvillian ${\cal{L}} : A (t) = \exp (i {\cal{L}} t )
A$, where $i {\cal{L}} A = (\partial A /
\partial Q) P / \mu - (\partial A /
\partial P)\mu \Omega^2 (\xi) Q$.

To embed the problem into the standard framework of
correlation--function theory a scalar product shall be introduced
in the space of variables $A, B \ldots$ by $(A \mid B) =
\overline{\langle A^* B \rangle}$. The vectors $\mid \! Q)$ and
$\mid \! P)$ are orthogonal and the normalization reads $(Q \mid
Q) = v^2 / \Omega^2 , \, (P \mid P) = \mu^2 v^2$. The Liouvillian
is hermitian. The displacement correlator shall be defined by
\begin{equation}
\label{Gl32}
\phi (t) = (Q (t) \mid Q) / (Q \mid Q) \,\, .
\end{equation}
Its Fourier--Laplace transform can be written as
Liouvillian--resolvent matrix element: $\phi (\omega) = (Q \! \mid
\! [{\cal{L}} - \omega]^{-1} \! \mid \! Q) / (Q \! \mid \! Q)$.
This quantity can now be represented within the Zwanzig--Mori
formalism as a double fraction \cite{Hansen86}: $\phi (\omega) = -
1 / [\omega - \Omega^2 / [\omega + \Omega^2 m (\omega)]]$. The
memory kernel $m (\omega)$ is the Fourier--Laplace transform of
the fluctuating--force correlator
\begin{mathletters}
\label{Gl33}
\begin{equation}
\label{Gl33a}
m (t) = (F (t) \mid F) / (v^2 \Omega^2 \mu^2) \,\, .
\end{equation}
Here $F$ is the projection of the force $\partial_t P = -\mu
\Omega^2 (\xi) Q$ perpendicular to $\mid Q)$ and $\mid P)$:
\begin{equation}
\label{Gl33b}
F = \mu \left[ \Omega^2 - \Omega^2 (\xi)\right] Q \,\, .
\end{equation}
The time evolution in Eq. (\ref{Gl33a}) is generated by the reduced
Liouvillian ${\cal{L}}^\prime , F(t) = \exp (i {\cal{L}}^\prime t)
F $, where ${\cal{L}}^\prime = {\cal{P L P}}$ and ${\cal{P}}$
denoting the projector perpendicular to $\mid \! Q)$ and $\mid \!
P)$. The susceptibility is connected with the correlator as
usual, $\chi (\omega) / \hat \chi = (\omega \phi (\omega) + 1)$,
\begin{equation}
\label{Gl33c}
\chi (\omega) = - \mu / \left[ \omega^2 - \Omega^2 + \omega
\Omega^2 m (\omega)\right] \,\, .
\end{equation}
\end{mathletters}
This exact representation of $\chi (\omega)$ in terms of kernel $m
(\omega)$ is the analogue of Eq. (\ref{Gl4d}).

The essential point in the MCT is the
approximation of the kernel $m (t)$ as mode--coupling functional.
The procedure \cite{Kawasaki70}, consists of two steps. 
Firstly, one reduces $F$
to the projection on the simplest modes contributing, and these
are the pair modes $\mid \! Q \cdot \xi) : F(t) \to \mid Q (t) \xi
) (Q \xi \mid Q \xi)^{-1} (Q \xi \mid F)$. Secondly, one
factorizes averages of products into products of averages: $(Q (t)
\xi \mid Q \xi) \to (Q (t) \mid Q) \overline{\xi^2}$. As a result
one finds
\begin{equation}
\label{Gl34}
m (t) = w_1 \phi (t)  \,\, , \qquad w_1 = \overline{ \xi \left[
(\Omega^2 / \Omega^2 (\xi) ) -1 \right]}^2 / \overline{\xi^2} \,\,
.
\end{equation}
These formulas are the analogue of Eqs. (\ref{Gl4}) and (\ref{Gl5}). 
They allow
the approximate evaluation of $\chi (\omega)$ from the given input
information $\Omega^2$ and $w_1$.

The Eqs. (\ref{Gl33c}) and (\ref{Gl34}) 
are equivalent to Eqs. (\ref{Gl19}) with
$\tilde \Omega = \Omega = \hat \Omega$, and therefore the approximation 
for $\chi (\omega)$ can be written as noted in Eqs. (\ref{Gl22}).
Thus, the presented MCT delivers an approximation for the
oscillator susceptitiblity (\ref{Gl31}) in the sense that the general
distribution $\rho (\xi)$ is approximated by $\tilde \rho (\xi) =
2 \sqrt{1 - \xi^2} / \pi$. If $\rho (\xi) = 2 \sqrt{1 - \xi^2} /
\pi$ is chosen, one can check that $\tilde \Omega = \Omega_0$ and
$w = w_1$. In this case MCT reproduces the exact
result. Let us add, that the naive factorization for
the kernel, $(F (t) \mid F) \simeq \mu^2 (Q (t) \mid Q)
\overline{[ \Omega^2 - \Omega (\xi)^2]^2}$, would not reproduce
the exact result for the specified example; rather one would
obtain Eq. (\ref{Gl34}) with an overestimated $w_1 = w + w^2$.

\section{Conclusions}

Within mode--coupling theory (MCT) a critical temperature $T_c$
and a corresponding critical packing fraction $\varphi_c$ were
introduced as the equilibrium--thermodynamics parameters
characterizing the evolution of glassy dynamics.
For silica, for example, $T_c$ is near 3300K \cite{Horbach99}, and
therefore all experiments quoted for this system in Sec.I deal
with $T \ll T_c$ states. In this paper primarily states are
studied where $T$ is so far below $T_c$ and $\varphi$ so far above
$\varphi_c$ that structural--relaxation phenomena do not dominate
the dynamics within the window of interest. These states are
referred to as stiff--glass states. The dynamical window
considered is the one of normal--condensed--matter physics, i.e.
spectra are discussed within a two--decade regime for the
frequency $\omega$ around and below the Debye frequency
$\omega_D$. For the stiff--glass states the spectra of the
$\alpha$--relaxation process are located at frequencies smaller
than $\omega_D / 100$ and therefore it does not matter whether or
not the quasielastic $\alpha$--peaks of the spectra are treated as
elastic ones. Hence it is legitimate to use the basic version of
the MCT which treats the crossover near $T_c$ as a sharp
transition to an ideal glass at $T_c$. The
derivation of the MCT formulas, in particular the one of Eq. (\ref{Gl2b})
for the mode--coupling coefficients, is based on
canonical--averaging properties. For temperatures below the
calorimetric glass--transition temperature $T_g$, the system is in
a quenched non--equilibrium state. From a rigorous point of view
the application of MCT is therefore restricted to the regime $T >
T_g$. However, experiments on high--frequency sound and on the
boson peak do not indicate anomalies for $T$ near $T_g$. Thus it
seems plausible that the results of the present paper can be used
also for an interpretation of $T < T_g$ data.

A major finding of this paper is that there are "peaks" of the
density--fluctuation spectra for wave vectors $q$ exceeding about
half of the Debye--vector $q_D$, which are quite different from
what one would expect for phonon resonances in liquids or
crystals. These peaks, which we refer to as the
anomalous-oscillation peaks (AOP), show the properties of the
so--called boson--peak listed in Sec.I. First, as shown in Fig.
7, the position $\omega_P$ of the peak maximum is several times
smaller than $\omega_D$. In this sense the AOP
is due to soft modes. Second, according to Eqs.
(\ref{Gl22}), the AOP is a superposition of harmonic--oscillator spectra,
where different oscillators are specified by different
frequencies. This can be shown since there is a well--defined
strong--coupling limit of the MCT equations, referred to as
stiff--glass limit, where the equations of motion simplify so much
that all features of the AOP can be worked out by analytical
calculations (Sec.IIIC). In this limit the continuous spectra are
purely inhomogeneous ones. Third, there is a lower cutoff
$\Omega_-$ of the frequency distribution. This causes the low
frequency wing of the AOP to decrease more steeply with decreasing
$\omega$ than expected for a Lorentzian. The high frequency wing
extends further out than the low frequency one so that the AOP is
skewed. Fourth, as the packing fraction decreases also the
frequencies $\omega_P$ and $\Omega_-$ decrease, and simultaneously
the intensity of the spectrum increases. This is shown in the
lower panel of Fig. \ref{fig_lin_ISF_w} and in Fig. \ref{fig_modulus}. 
The critical point is
characterized by the threshold $\Omega_-$ approaching zero. In
this sense one concludes, that the evolution of the AOP is related
to the dynamical instability predicted by MCT for $\varphi =
\varphi_c$ or $T = T_c$. In order to understand a further property
of the AOP one has to acknowledge that the leading correction to
the stiff--glass results introduces a damping for the oscillators.
It is due to the decay of an oscillator mode into two modes caused
by anharmonicities. A simplified treatment of this phenomenon only
leads to a modification of the formulas by the introduction of a
friction constant $\nu$ in Eq. (\ref{Gl22c}). Thus the peak is
superimposed on a flat background and the sharp thresholds are
changed to some smooth but rapid crossover. Not much is modified
in the center of the AOP spectrum provided $\nu$ is not too big.
But with decreasing $\varphi$ or increasing $T$, the ratio $\nu /
\Omega_-$ becomes much larger than unity so that the oscillators
of low frequency get overdamped. As a result one obtains the
explanation of the fifth property, namely a central relaxation
peak is formed if $(\varphi - \varphi_c) / \varphi_c$ is about
10\%. In this case the AOP merely appears as a shoulder of the
quasielastic spectrum as shown by the $n = 4$ curve in the lower
panel of Fig. \ref{fig_lin_ISF_w}. Shifting the parameters even closer 
to the
instability point, the AOP gets buried under the wing of the
quasielastic peak. The elementary formulas (\ref{Gl19}) and (\ref{Gl20}) 
describe
this feature of the MCT solutions reasonably well for $\varphi$
approaching $\varphi_c$ up to about 5\%. The cited results are
quite general and are obtained even for the simplest schematic MCT
models, as is demonstrated in Fig. \ref{fig_F12_deep}. Sixth, in a further
refinement of the description one acknowledges that the glass
compressibility has a wave--vector dependence. This enters in the
form of a characteristic frequency $\hat \Omega_q$, which can be
considered as a bare phonon dispersion of the amourphous solid. It
exhibits a maximum for $q$ near $q_D$ and a
minimum near the structure--factor--peak position $q_{\max}$ which
in turn is near $2 q_D$. This $\hat \Omega_q$--versus--$q$ curve
is similar to the one for the characteristic frequency $\Omega_q$
of the liquid which plays an essential role in the MCT equations
(\ref{Gl1a}) and (\ref{Gl1b}). However, the oscillations of the 
Debye--Waller
factor, shown in Fig. \ref{fig_DWF}, imply via the renormalization 
formula
(\ref{Gl4b}), that the ratio of the maximum to the minimum frequency is
much smaller for $\hat \Omega_q$ than for $\Omega_q$. This can be
inferred in detail by comparing the full line in Fig. \ref{fig_Omega} 
with the
dashed one. Therefore, the maximum position of the AOP is only
weakly $q$--dependent. Summarizing, we suggest that the MCT of the
AOP provides the basis of a first--principle explanation of the
so--called boson peak.

The formulated theory for the AOP has a transparent
interpretation. MCT explains in the first place the formation of
an effectively arrested density distribution. This
amorphous structure is characterized by the same quantity used to
characterize crystalline structures, viz. by the Debye--Waller
factor $f_q$. Within the same formalism which leads to $f_q$, the
equations of motion for the density fluctuations of this structure
are obtained (Sec.IIB). Such unified treatment of the glass
structure and its dynamics should be a feature of every
microscopic theory since it is the same array of particles which
forms the frozen structure and which carries the fluctuations.
This unified treatment is especially important if one intends to
study the dynamics near the instability limit of the structure.
The equations of motion describe the decay of fluctuations into
pairs and also the scattering of fluctuations from the arrested
structure. For the stiff--glass states one finds the latter
processes to overwhelm the former, as discussed in connection with 
Eq. (\ref{Gl12}). In the stiff--glass limit one
finds that the particles are localized in their cages and harmonic
oscillations of the particles with their cages are a good
description of the relevant modes. In this extreme limit the total
susceptibility is the one of a distribution of
independent--oscillator responses, where the distribution of
oscillator frequencies is caused by the distribution of sizes and
shapes of the cages. MCT provides an approximation of the
distribution of the frequency squares, Eq. (\ref{Gl22b}), and
characterizes the drift of the distribution with changes of
control parameters. It provides also results for the corrections
to this limiting result, namely the appearance of homogeneous line
broadening due to mode decay and coupling of the oscillations
leading to weak wave--vector dependencies of the AOP parameters.

One achievement of MCT is the possibility to explain homogeneous
line--broadening effects via a golden--rule mechanism as is
suggested by Eqs. (\ref{Gl5d}) and (\ref{Gl5e}). This aspect was used 
above in
connection with the evaluation of the oscillator damping $\nu$ due
to two--mode decay and also in connection with the derivation of
Rayleigh's law, Eq. (\ref{Gl15a}). However, the interpretation of Eqs.
(\ref{Gl5d}) and (\ref{Gl5e}) in the spirit of a golden rule is quite 
misleading,
if the transition probabilities $\hat V$ are so large that the
self--consistent solutions are qualitatively different from the
ones obtained by a lowest--order approximation. In such cases the
formulas can lead to an approximation theory for an inhomogeneous
"line width" phenomenon. It was shown explicitly in Sec.IVB that
MCT provides an approximation approach towards this phenomenon,
and there is an example for which MCT reproduces the exact result
for the inhomogeneous spectrum. The MCT for the AOP is based on
the fact that this theory can handle homogeneous and inhomogeneous
spectra within the same framework.

A side remark might be helpful. According to MCT all structural
relaxation features are independent of the details of the
microscopic equations of motion \cite{Goetze92,Franosch98}.
Therefore the existence of an AOP is of no relevance for
understanding glassy dynamics or the glass transition. But the AOP
provides an interesting piece of information on the arrested glass
structure. The AOP is the result of a mapping of the cage distribution
on the frequency axis. The mapping is done from the configuration
space on the time axis via Newton's equations of motion and
canonical averaging followed by a Fourier--cosine transform to get
a spectrum.

The arrest of density fluctuations at the ideal liquid--glass
transition is driven by the ones with a wave number $q$ near the
structure--factor--peak position $q_{\max} \approx 2 q_D$, since
for these wave numbers the liquid compressibility $\chi_q^T
\propto S_q$ is large. The compressibility of simple dense liquids
for $q < q_D$ is very small and therefore excitations with wave
vectors from this domain, which corresponds to the first Brillouin
zone of the crystalline phase, are not important for the evolution
of structural relaxation and the glass transition. For the same
reason one concludes that scattering processes of density
fluctuations with $q < q_D$ are irrelevant for the formation of
the AOP, as is demonstrated in Fig. \ref{fig_V0}. The soft complexes 
which
cause the AOP are constructed from fluctuations with wave numbers
near and above $q_{\max}$. These conclusions are based on the MCT
results for the Debye--Waller factors $f_q$. In order to produce
the spectral peak in Fig. \ref{fig_scat_cont}, $f_q$ has
to be that large as shown by the uppermost curve in Fig. \ref{fig_DWF}. 
This
curve corresponds to $\epsilon = (\varphi - \varphi_c) / \varphi_c
= 0.16$. In Fig. \ref{fig_HSM_ISF_w} of Ref. \cite{Megen95} a 
measurement of $f_q$
for a hard--sphere colloid is documented for $3 \leq q d \leq 13$
and $\epsilon = 0.11$. Since these experimental findings are close
to the curve (a) in Fig. \ref{fig_DWF} we argue that the MCT results 
on the
glass structure are in reasonable accord with the experimental
facts. Let us emphasize, that the above reasoning refers to
densely--packed systems of spherical particles. Obviously, in more
complicated systems, like silica, the cages are not so tight like
in a HSS. Therefore one can expect the soft configurations to be
more subtle than discussed in this paper. The results in Sec.IVA
for the two--component schematic model indicate, that the AOP can
be more structured than obtained for the HSS. For systems with a
low coordination number one can also expect
intermediate range--order effects to play an important role. They
enter, e.g., the coupling vertices in Eq. (\ref{Gl2b}) via the 
prepeaks of
$S_q$ \cite{Foley95}. Whether MCT can handle the microscopic
features leading to the AOP in complicated systems such as 
ZnCl$_2$ is unclear at present. In particular, it is unclear 
whether MCT can contribute to understanding why the boson peak is more 
pronounced in strong glass formers like silica than in fragile glass 
formers like orthoterphenyl.

The preceding interpretation of the AOP suggests that these peaks
appear in the spectra of all probing variables coupling to density
fluctuations of short wave length. But different probing variables
will weight the oscillating complexes differently and therefore
the shape of the AOP and the position $\omega_P$ of the peak
maximum will depend somewhat on the probe. Let us consider two examples. 
The first one deals with the tagged
particle correlator $\phi_q^s (t) = \langle \rho_{\vec
q}^s(t)\rho_{\vec q}^{s*} \rangle$. Here $\rho_{\vec q}^s (t) =
\exp (i \vec q \vec r (t))$ denotes the density fluctuation of a
marked particle with position vector $\vec r (t)$. The spectrum
$\phi_q^{s \biss} (\omega)$ determines the incoherent
neutron--scattering cross section. An exact equation
for this quantity has the same form as Eq. (\ref{Gl1a}) with 
$\phi_q, m_q$
and $\Omega_q^2$ replaced by $\phi_q^s, m_q^s$ and $\Omega_q^{s2}
= v^2 q^2$, respectively. The essential MCT equation is again the
representation of the kernel as mode--coupling functional $m_q^s
(t) = \sum_{kp} V_{q, kp}^s \phi_k (t) \phi_p^s (t)$. The coupling
coefficients $V_{q, kp}^s$ are determined by the structure factor
$S_q$. The mentioned equations have been derived and solved for
the Lamb--M{\"o}ssbauer factors $f_q^s = \phi_q^s (t \to \infty)$
of the HSS in Ref. \cite{Bengtzelius84}. Details of the
discretization can be found in Ref. \cite{Fuchs98}. We have
solved the cited equations for $\phi_q^s (t)$, and Fig. \ref{linsisfw}
exhibits fluctuation spectra for the packing fraction $\varphi =
0.60$ for three wave vectors around the structure--factor--peak
position. The shape of the peaks is only weakly $q$--dependent and
the intensity varies nearly proportional to $q^2$. This finding is
in agreement with neutron--scattering results \cite{Cusack90} and
with molecular--dynamics--simulation results 
for ZnCl$_2$ \cite{Foley95} and silica \cite{Horbach99c} .

The second example is related to the velocity correlator $\Psi (t)
= \langle \dot{\vec r} (t) \dot{\vec r} \rangle / (3 v^2)$. For a
harmonic system its spectrum determines the density of states $g
(\omega) = 2 \Psi^{\biss}(\omega) / \pi$, normalized by $\int_0^\infty g
(\omega) d\omega = 1$. The velocity correlator can be extracted
from $\phi_q^s$ \cite{Hansen86}. If one introduces the kernel
$m^{(0)} (t) = \lim_{q \to 0} q^2 m_q^s (t)$ one gets $\Psi
(\omega) = - 1 /[\omega + v^2 m^{(0)} (\omega)]$. The localization
length $r_s^2$, defined via the long time limit of the mean
squared displacement $\langle \delta r^2 (t \to \infty) \rangle =
6 r_s^2$, determines the small--$q$ asymptote of $f_q^s$ and the
pole of the kernel in the glass: $f_q^s = 1 - (q r_s)^2 + O (q^4),
\, m^{(0)} (\omega) = - 1 / (\omega r_s^2) + O (\omega^0)$. One
finds that the density of states vanishes proportional to
$\omega^2$ for small frequencies as expected for an elastic
continuum $g (\omega) = g_0 \cdot \omega^2 + O (\omega^4) , \, g_0
=  m^{(0) \biss} (\omega = 0) (2 r_s^4 / \pi v^2)$
\cite{Goetze91b}. Figure \ref{fig_dos} exhibits a result for the HSS. 
The
$\omega^2$--law is obtained only for frequencies below the
threshold $\Omega_-$ of the AOP. For larger frequencies the
density of states is enhanced relative to the asymptotic law. The
enhancement reaches a maximum of a factor of about 5 near the
position of the AOP maximum, and it is about a factor 3 at the
maximum of $g(\omega)$. For larger frequencies $g (\omega)$ gets
suppressed as required by the normalization condition. The found
enhancement phenomenon is in qualitative agreement with the
experimental results reported for silica in Ref. \cite{Buchenau86}
and with the simulation data in Ref. \cite{Taraskin99b}. There is
a consistency problem for the MCT. The prefactor in the
$\omega^2$--law should be $g_0^\prime = \omega_D^{-3} [1 + 2
(\omega_D / \omega_D^\prime)^3]$, where $\omega_D^\prime$ denotes
the Debye--frequency for transversal sound. Since $g_0$ was
calculated without any explicit reference to transversal sound,
the approximations underlying MCT will lead to $g_0 \not=
g_0^\prime$. We did not study the solutions of the MCT equations
for transversal excitations \cite{Bengtzelius84,Goetze91b} in
order to calculate $\omega_D^\prime$. Therefore we do not know the
size of the error $g_0 - g_0^\prime$. But if one estimates
$\omega_D = \omega_D^\prime$ one gets $g_0^\prime = 10.2 \,\,\,
10^{-8}$, which is close to the value $g_0 = 9.2 \,\,\, 10^{-8}$.
This suggests that transversal excitations are taken into account
to some extend implicitly in the formulas for $m_q^s (t)$.

It was already suggested earlier in connection with a discussion
of soft--configuration models for glasses that the boson peak
should be understood as result of quasi--harmonic oscillations of
the system characterized by some distribution of oscillator
potentials \cite{Karpov83,Klinger88,Galperin85,Buchenau92}.
Obviously, the present theory is consistent with these
phenomenological approaches. In Ref. \cite{Sokolov94} Raman
spectra of glassy systems have been interpreted as a superposition
of oscillator susceptibilities analogous to what is formulated in
Eqs. (\ref{Gl22}). But there are two qualitative differences between 
this
fit procedure and the present theory. In Ref. \cite{Sokolov94} the
distribution $\tilde \rho (\xi)$ is taken as temperature
independent, while Eq. (\ref{Gl22b}) for $\tilde \rho (\xi)$ describes 
the
softening of the glass structure upon heating and, in particular,
its instability for $T$ reaching $T_c$. In Ref. \cite{Sokolov94}
the $T$ dependence of the spectra is introduced by replacing the
damping constant $\nu$ by a Debye--function quantified by a
temperature--dependent relaxation time. This visco--elastic
theory leads to a Debye peak as
quasi--elastic spectrum. Equations (\ref{Gl22}) do not lead to a
quasi--elastic Debye spectrum as was explained in
connection with Figs. \ref{fig_HSM_ISF_w} and \ref{fig_lin_ISF_w}. 
In Refs.
\cite{Schirmacher89,Schirmacher92} the effective--medium theory
for percolation problems is modified to a theory for the
displacement susceptibilities of a disordered harmonic lattice.
For the susceptibility an expression similar to Eq. (\ref{Gl16}) is
obtained where kernel $K^{SGA} (\omega)$ also describes the
self--consistent treatment of phonon scattering by the disorder.
Even though the equation for $K^{SGA} (\omega)$ in Refs.
\cite{Schirmacher89,Schirmacher92} is quite different from Eq.
(\ref{Gl17}), the solution looks similar to the dashed line in 
Fig. \ref{fig_mz}.
However, the Rayleigh contribution in Ref. \cite{Schirmacher92} is
about $10^6$ times larger than the result based on Eq. (\ref{Gl18}). It
was criticized \cite{Buchenau92} that in Refs.
\cite{Schirmacher89,Schirmacher92} the boson peak is constructed
from fluctuations with wave vectors $q < q_D$ since thereby 
the role
of long--wave--length fluctuations is overestimated. 
Hence the cited harmonic--lattice
theory does not appear to be compatible with the 
theory studied in this paper.

Sound is obtained due to the interplay of inertia effects and
stresses, which are built up due to compressions. The interplay is
governed by the conservation laws for mass and momentum. The
low--lying sound excitations interfere with other low--lying modes
like structural relaxation and the oscillations building the AOP.
Sound modes with wave vector $q \leq q_D$ and their interactions
are not important for the explanations of structural relaxation
\cite{Bengtzelius84}. In connection with Fig. \ref{fig_V0} it was shown,
that they are neither relevant for the explanation of the AOP.
Therefore sound can be discussed within the standard procedure of
acoustics by introducing a modulus. 
This procedure comes out within MCT as was explained in
connection with Eq. (\ref{Gl9}), where $[1 - \omega K(\omega)]$ is
proportional to the complex modulus. Consequently, all results on
high--frequency sound, discussed in this paper, are implications
of the preceding results for $K(\omega)$. In this sense one
concludes that high--frequency sound is a manifestation of the
AOP. Let us contemplate the scales for frequency $\omega$ and wave
number $q$ in order to be able to correlate the MCT results with
some data. The boson--peak maximum observed for silica at 1 THz
\cite{Foret96} shall be compared with the maximum position
$\omega_P = 75$ for the $\varphi = 0.60$ results for the AOP. The
best resolution width $\Gamma_{\exp}$ achieved by the recent
X--ray--scattering experiments is 1.5 meV, i.e. in the units of
this paper $\Gamma_{\exp} \approx 30$. The structure factor peak
position $q_{\max}$ = 1.5 \AA$^{-1}$ for silica is to be compared
with the value near 7 for the HSS. Hence the wave vector unit used
is about $2 nm^{-1}$. Scattering experiments with the resolution
$\Gamma_{\exp}$ have been done for silica for $q$ between $1
nm^{-1}$ and $4 nm^{-1}$ as can be inferred from Ref.
\cite{Masciovecchio97} and the papers quoted there.
X--ray--scattering experiments with larger $q$ are done with a
resolution considerably worse than the cited $\Gamma_{\exp}$. Thus
the following discussion shall be restricted to wave vectors
between $q = 0.5$ and $q = 2 \approx q_D /2$. This wave--vector
interval corresponds to the interval for the sound frequency $\hat
\Omega_q = v_\infty q$ between about 40 and 150 as is shown in 
Fig. \ref{fig_Omega}.

The first property of high--frequency sound follows from the upper
panel of Fig. \ref{fig_modulus}. On the resolution scale 
$\Gamma_{\exp}$ the
sound--dispersion law is $\Omega_q^{\max} = \hat \Omega_q =
v_\infty q$. Here the sound speed $v_\infty$ is the one determined
by the $q \to 0$ limit of the glass compressibility $\hat \chi_q$.
The frequency--dependence of the reactive part of the modulus $1 -
\omega K^\prime (\omega)$ implies deviations from the strict
linear law for $\Omega_q^{\max}$. According to Fig. \ref{fig_Omega} 
the deviations are predicted to occur on a 10\% level. Thus they
should be measurable if the resolution $\Gamma_{\exp}$ could be
reduced by, say, a factor 5. Here some reservation has to be made.
MCT does not contribute to the understanding of the structure
factor $S_q$, rather it takes this quantity from other theories.
Errors in $S_q$ will cause errors in the MCT results. It is
notoriously difficult to calculate the small--$q$ behavior of
$S_q$. The Verlet--Weiss theory yields different results for $S_q$
than the used Percus--Yevick theory; and this causes also a
small--$q$ behavior of $f_q$ which differs from the one exhibited
in Fig. \ref{fig_DWF} \cite{Fuchs92c}. Therefore the Verlet--Weiss 
theory will
also lead to different $\hat \Omega_q$ which might change the
results for $\Omega_q^{\max}$. It is not known how reliably the
Percus--Yevick or the Verlet--Weiss theory describe the structure
for packing fractions as large as 0.60.

If one smears out the spectrum $K^{\biss} (\omega)$ shown in Fig.
10 with a resolution curve of width $\Gamma_{\exp}$, one gets a
result which is nearly $\omega$--independent within the dynamical
window of interest: $v_\infty^2 K^{\biss}(\omega) = \gamma$. This
explains the second property of high--frequency sound reported for
the X--ray--scattering results of silica
\cite{Benassi96,Masciovecchio97} and other systems
\cite{Masciovecchio96,Monaco98,Matic99}: the sound damping
exhibits the hydrodynamic wave vector dependence $\Gamma_q =
\gamma q^2$. Consequently, one can describe the whole measured
spectrum by the damped--oscillator formulas (\ref{Gl6a}) and (\ref{Gl6b})
\cite{Masciovecchio96,Masciovecchio98,Masciovecchio97}, albeit up
to some background. The latter appears as white if viewed with
resolution $\Gamma_{\exp}$. Naturally, it is difficult to separate
this background from the one caused by other effects of the
experimental setup. If one acknowledges the frequency dependence
of $K (\omega)$, exhibited in Fig. \ref{fig_modulus}, one concludes, 
that the
formula $\Gamma_q \propto q^2$ is oversimplified. Figure \ref{fig_Gamma}
demonstrates the prediction, that a reduction of $\Gamma_{\exp}$
by a factor of three should be sufficient to detect an increase of
$\Gamma_q$ above the $\Gamma_q \propto \Omega_q^{\max 2}$
asymptote if $q$ varies between 0.8 and 2.

A crucial experimental finding is that the damping parameter
$\Gamma_q$ does not vary much with changes of temperatures. This
shows \cite{Benassi96,Monaco98}, that the sound damping mechanism
cannot be due to anharmonicity--induced mode decay as known for
phonons in crystals nor due to structural relaxation effects as
studied in Brillouin--scattering spectroscopy of glassy liquids.
The fast sound detected by neutron scattering
\cite{Teixeira85} and molecular-dynamics simulation \cite{Sciortino94} 
in water, for example, occurs in a dynamical window where water 
exhibits its $\alpha$--relaxation process but no vibrations
underlying a boson peak \cite{Sokolov95,Sciortino97}. Therefore 
the fast--sound-- damping in water depends appreciably on 
temperature and this dependence can be described reasonably 
within a visco--elastic model \cite{Sciortino94}. The insensitivity 
of the high--frequecy--sound damping on changes of control--parameters 
like temperature or density is indeed the third specification 
of the MCT results as shown in Fig. \ref{fig_fast_sound}. 
In agreement with the assumptions of the
phenomenological theories in Ref. \cite{Klinger88,Buchenau92} MCT 
explains the damping to be due to absorption of the sound mode by
the oscillations building the AOP. The AOP depends on control
parameters, as explained in connection with Figs. \ref{fig_modulus} and 
\ref{fig_F12_deep}.
Therefore $\Gamma_q$ is not strictly independent of control parameters.
However, changes of density or temperature primarily redistribute
the spectrum of $K^{\biss}$ and thus the spectrum in the centre of
the AOP does not change much. But, if the resolution
$\Gamma_{\exp}$ could be reduced, a more subtle prediction could
be tested. For $\Omega_q^{\max}$ near 140, $\Gamma_q$ is
$\varphi$--independent even if the system is driven as close to
the critical point as shown by curve n = 4 in Fig. \ref{fig_modulus}. 
For smaller
$\Omega_q^{\max}$ the width $\Gamma_q$ increases with decreasing
$\varphi$, and this is due to the appearance of the quasi--elastic
relaxation peak of $K^{\biss}(\omega)$. But for larger
$\Omega_q^{\max}$, the damping constant should decrease with
decreasing $\varphi$; in this case the softening of the system
reduces the density of states for high--frequency--sound--decay
processes. Let us emphasize that all MCT results discussed in this 
paper are solely based on the wave--vector and control--parameter 
dependence on the structure factor $S_q$. The explanation of the 
modulus and its drift with control-parameter changes 
is specified semi 
quantitatively by the three numbers only, which are specified in 
connection with Eqs. (\ref{Gl20}) and (\ref{Gl21}). Therefore our 
results are predicted to be valid for all systems with a structure 
factor similar to that of our HSS model, in particular for 
Lennard--Jones systems or van der Waals 
liquids like, e.g., orthoterphenyl.

The AOP of $K^{\biss}(\omega)$ causes via Kramers--Kronig
relations a frequency dependent reactive part $K^\prime (\omega)$.
This implies that the sound resonance cannot exhaust the spectrum.
Contrary to what holds for hydrodynamic sound, there must be a
background spectrum. This is the fourth feature specified for
high--frequency sound in Sec.I. Obviously, the detection of such
background is difficult in view of many other reasons producing
backgrounds for the experimental scattering signals. However, as
explained in connection with the dashed lines in Figs. \ref{fig_mz} 
and \ref{fig_F12_deep},
MCT implies a fifth property: for $T \ll T_c$ or
$\varphi \gg \varphi_c$, there is an effective low--frequency
threshold $\Omega_-$ for the background. Such threshold can be
used to discriminate the background due to density--fluctuation
dynamics from the one due to experimental artifacts. There is a
mathematically equivalent manner to formulate the physics of the
background, which is better adopted to the present problem
\cite{Sampoli98,Taraskin99}. The dynamical structure factor $S (q,
\omega) = S_q \phi_q^{\biss} (\omega)$, considered as function of
$q$ for fixed frequency $\omega$, represents the average of the
square of the density--fluctuation Fourier components which
oscillate with frequency $\omega$. The coherent contribution to
these fluctuations leads to a peak at $q_\omega = \omega /
v_\infty$. There is only a small contribution for $q < q_\omega$,
since it is very difficult to excite long--wave--length
fluctuations in densely packed systems. But there is a
structureless background for $q > q_\omega$ extending to high
values of $q$. It is caused by the large--$q$ density fluctuations
produced by the distortions of the wave front due to the arrested
amorphous glass structure. Figure \ref{fig_Sqw} exhibits a MCT result 
for the
HSS, which is in qualitative agreement with the simulation results
reported in Ref. \cite{Sampoli98}.

All specified MCT results for high--frequency sound and the AOP can be
described well by the combination of the elementary formulas (\ref{Gl9})
and (\ref{Gl20}) with $K (\omega) \approx \tilde K (\omega)$. If regimes
are considered, where relaxation can be ignored completely, one
can use Eq. (\ref{Gl20b}) with $\nu = 0$. In this case, only the two
parameters $\tilde \Omega$ and $w_1$ need to be specified in order
to quantify the result. Function $\tilde K (\omega)$ replaces the
damping parameter $\gamma$ of the hydrodynamics--theory 
result, Eqs. (\ref{Gl6}). Introducing the
third parameter $\nu$, the range of applicability of the results
can be extended so, that structural--relaxation precursors are
included. We suggest to use the cited formulas for an analysis of
inelastic--X--ray--scattering data for high--frequency sound and
of data for the evolution of the boson peak in glassy systems.

The derivation of the MCT equations is definitive and
leads to a well--defined model for a non--linear
dynamics. The point of view adopted in this paper is the
following: results for the model are derived to provide
explanations of previously unexplained features of the dynamics of
liquids and glasses and to predict new results to be tested by
future experiments. However, the "approximations" leading to the
mode--coupling expression for the fluctuating--force kernel, Eqs.
(\ref{Gl2}), are uncontrolled and therefore the range of validity of MCT
is not understood. Let us conclude this paper by pointing out four
open questions concerning the foundation of MCT, which are of
particular relevance for the study of vibrational excitations. The
first problem concerns the absence of any influence of transversal
excitations on the dynamical structure factor. The results (\ref{Gl15a})
and (\ref{Gl18}) for Rayleigh's law account for the scattering of a
longitudinal phonon into some other longitudinal wave, while the
expected contribution due to conversion into transversal sound
waves is missing. Similarly, the spectrum of the AOP is due to
longitudinal excitations only, while Horbach et al.
\cite{Horbach99b} report that transversal modes
influence the boson--peak. The second problem concerns
the mode--coupling approximation for the
fluctuating--force kernel $m_q (t)$ for short times. For a
Lennard--Jones system Eqs. (\ref{Gl2a}) and (\ref{Gl2b}) yield an 
overestimation
of $m_q (t = 0)$ compared to the values known from Monte--Carlo
results. A procedure for eliminating this problem was suggested in
Ref. \cite{Sjoegren80}, but it is unclear whether it 
can be used for supercooled systems. One also expects that the
mode--coupling kernel should be complemented by some regular term
\cite{Bengtzelius84}. In a simple treatment this would lead to an
additional friction term $\nu_q \dot \phi_q (t)$ in Eq. (\ref{Gl1a}). 
For a dilute system $\nu_q$ could be calculated in the
binary--collision approximation. The third unsolved problem is the
evaluation of $\nu_q$ for the dense systems under consideration.
Such friction term would imply an additive correction to the
friction constant $\nu$ in Eqs. (\ref{Gl20}) and (\ref{Gl22}). 
If the resulting $\nu$ would be too big, the AOP could disappear 
in favour of an quasi--elastic spectrum due to overdamped
oscillations. Whether this happens might depend on the
structure of the system. As fourth unsettled question a cut--off
problem has to be mentioned. For the large packing fraction
$\varphi = 0.60$, the results of our
calculations would change somewhat if the used cutoff wave vector
$q^* d =40$ was increased. This is due to the slow decrease
towards zero of the direct correlation function $c_q$ of the HSS
for $q$ tending to infinity. Since the introduction of a cutoff is
equivalent to a softening of the hard--sphere potential it is
conceivable that in the stiff--glass limit our model is rather a
model for argon than for hard spheres. It would be desirable to
examine whether this cutoff dependence disappears if a
conventional regular interaction potential is used. A serious
bottleneck for such examination is the necessity to obtain
reliable results for $c_q$ for strongly supercooled liquids.

\bigskip
\bigskip

\noindent {\bf Acknowledgments}

We thank H.Z. Cummins, J. Horbach, A. Latz and W. Schirmacher for
helpful discussions and critical comments on the manuscript.

\newpage

\begin{figure}
\caption{Debye--Waller factor $f_{q}$ (crosses) and one tenth of
the Percus--Yevick static structure factor $S_{q}$ (lines) of a
hard--sphere system (HSS) at the packing fractions $\varphi=0.600$
(a), $\varphi= 0.540$ (b), and the critical packing fraction
$\varphi=0.516$ (c). 
The inverse of the particle diameter, $1/d$, is chosen as unit 
for the wave vector $q$. }
\label{fig_DWF}
\end{figure}

\begin{figure}
\caption{Density correlation functions $\Phi_{q}(t)$ of a HSS as a
function of time $t$ for the wave numbers $q=3.4$ and $q=7.0$. The
curves refer to the packing fractions
$\varphi=0.6$ and $\varphi=\varphi_{c}(1\pm10^{-n/3})$ with $n$
given in the figure. Here $\varphi_{c}\approx0.516$ denotes the
critical packing fraction. The curves with label $c$ are the
solutions at the critical
point, which approach the long--time limits $f_{3.4}^{c}=0.36$ and 
$f_{7.0}^{c}=0.85$. The units of length and time have been chosen 
here and in all the following figures
so that the hard--sphere diameter $d=1$ and the thermal 
velocity $v=2.5$. }
\label{fig_HSM_ISF_t}
\end{figure}

\begin{figure}
\caption{Fluctuation spectra $\Phi_{q}^{\prime\prime}(\omega)$ of
the
correlation functions shown in Fig. \ref{fig_HSM_ISF_t}. }
\label{fig_HSM_ISF_w}
\end{figure}

\begin{figure}
\caption{Some of the correlators from Fig. \ref{fig_HSM_ISF_t} on
a linear time axis. The full lines refer to glass states $(n = 3 :
\varphi = 0.567; \, n = 4 : 0.540)$ with the arrows indicating the
long--time limits $f_{q}=\lim_{t\rightarrow0}\Phi_{q}(t)$. The
dashed lines with label $c$ exhibit the critical decay $(\varphi =
0.516)$ and the lowest dashed curves refer to the liquid state for
$n=4$ $(\varphi = 0.549)$.
}
\label{fig_lin_ISF_t}
\end{figure}

\begin{figure}
\caption{Fluctuation spectra $\Phi_{q}^{\prime\prime}(\omega)$ of
the correlators shown in
Fig. \ref{fig_lin_ISF_t}. }
\label{fig_lin_ISF_w}
\end{figure}

\begin{figure}
\caption{Spectra $\Phi_{q}^{\prime\prime}(\omega)$ (solid lines)
of a HSS at packing fraction $\varphi=0.60$ for some wave numbers
$q$. The dashed lines show the generalized--hydrodynamics
approximation. For $q=1.0$ and $q=3.4$ the dotted lines show the
generalized--hydrodynamics approximation with $f_q$ replaced by
$f_{q=0}$.
}
\label{fig_happrox}
\end{figure}

\begin{figure}
\caption{Frequency $\Omega_{q}=vq/\sqrt{S_{q}}$ (dashed line),
renormalized frequency
$\hat{\Omega}_{q}=\Omega_{q}/\sqrt{1-f_{q}}$ (full line) and
position of the global maximum $\Omega_{q}^{\max}$ of the spectrum
(diamonds) of a HSS at packing fraction $\varphi=0.60$ as a
function of the wave number $q$. For wave numbers at which two
separate maxima of the spectrum can be identified the frequency
position of the peak with the lower intensity is marked by the
open circles. The vertical bars mark the frequency intervals where
$\Phi_q''(\omega)$ exceeds half of the maximum intensity of the
spectrum. The arrows point to the positions of the Debye wave
number $q_D=(36\pi\varphi)^{1/3}=4.08$ and the Debye frequency
$\omega_D=v_\infty q_D=309$ corresponding to the high--frequency
sound speed $v_\infty=75.8$. For $q=3.4$ ($q=7.0$) one gets
$\Omega_{3.4}=77.0$ ($\Omega_{7.0}=13.7$) and
$\hat{\Omega}_{3.4}=167$ ($\hat{\Omega}_{7.0}=92.3$).
}
\label{fig_Omega}
\end{figure}

\begin{figure}
\caption{Width at half maximum $\Gamma_{q}$ (diamonds) of the
high--frequency resonance of the spectrum
$\Phi_{q}^{\prime\prime}(\omega)$ of the HSS for packing fraction
$\varphi=0.60$ as a function of the resonance--maximum position
$\Omega_{q}^{\max}$ for various wave vectors. The straight line
represents the small--wave--vector asymptote
$\Gamma_q=\gamma(\Omega_q^{max}/v_\infty)^2=K''(\omega=0)(\Omega_q^{max})^2$
where $K''(\omega=0)=0.00182$.
}
\label{fig_Gamma}
\end{figure}

\begin{figure}
\caption{High--frequency sound--resonance position $\Omega_q^{max}$
(diamonds)  and resonance width $\Gamma_q$ (circles) as a function
of the packing fraction, determined for $q=1.8$. The crosses
exhibit the maximum position $\omega_P$ of the AOP of the
density--fluctuation spectrum for wave vector $q=7.0$.
}
\label{fig_fast_sound}
\end{figure}

\begin{figure}
\caption{Reactive parts of the moduli $\Delta(\omega)=1-\omega
K'(\omega)$ and spectra $K''(\omega)$ of the HSS for the packing
fractions $\varphi=0.600$ (solid), $\varphi=0.567$ ($n=3$,
dashed) and $\varphi=0.540$ ($n=4$, dotted).
}
\label{fig_modulus}
\end{figure}

\begin{figure}
\caption{Mode--coupling coefficients $W_{q}^{(1)}$ and
$W_{q}^{(2)}$ determining via Eqs. (\ref{Gl10}) the scattering and decay
contributions, respectively, to the kernel $K$ of the HSS glass
states. The curves with label $n=3$ and $n=4$ refer to the packing
fractions $\varphi=0.567$ and $\varphi=0.540$, respectively. The
insets show the coefficients for $q<3$ magnified by a factor 100.
}
\label{fig_V0}
\end{figure}

\begin{figure}
\caption{The spectrum $K''(\omega)$ of the kernel for the HSS at
packing fraction $\varphi=0.60$ reproduced from Fig.
\ref{fig_modulus} (full line). The dashed and the dotted curves
exhibit the one--mode contribution $K^{(1)\prime\prime} (\omega)$
and the two--mode contribution $K^{(2)\prime\prime} (\omega)$,
respectively. The full line with label $R$ denotes the Rayleigh
contribution $R_0\omega^2$ ($R_0=1.5 \  10^{-11}$) magnified by a
factor $10^{4}$.
}
\label{fig_scat_cont}
\end{figure}

\begin{figure}
\caption{The full line reproduces the spectrum $K''(\omega)$ from
Fig. \ref{fig_scat_cont}. The dashed line exhibits the
stiff--glass--approximation result $\tilde{K}''(\omega)$ from Eqs.
(\ref{Gl20}) with $\tilde{\Omega}=120$, $w_1=0.388$, $\nu=0$. The dotted
line is the extension of this approximation by incorporating
$\nu=\tilde{\Omega}^2\tilde{K}^{(2)\prime\prime}(\omega=0) \approx
23$. The dot--dashed line shows the extension with $i\nu$ replaced
by the kernel $\tilde{\Omega}^2\tilde{K}^{(2)}(\omega)$ from 
Eq. (\ref{Gl21}).
}
\label{fig_mz}
\end{figure}

\begin{figure}
\caption{Fluctuation spectra $\Phi^{\prime\prime}(\omega)$ for the
one--component schematic model defined in Sec.IVA for a 
mode--coupling functional ${\mathcal{F}}[f]=v_1f+v_2f^2$ (solid
lines) and $\Omega=1$. The states refer to the glass with distance
parameters $\epsilon=1/4^n$, $n=0,1,2$ (compare text). 
The dashed line exhibit the stiff--glass approximation 
given by Eqs. (\ref{Gl20}) with $\nu=0$. The
dotted lines show the extended description including a $\nu\not=0$
which was evaluated from the $\omega=0$ limit of Eq. (\ref{Gl21}).
}
\label{fig_F12_deep}
\end{figure}

\begin{figure}
\caption{Spectra $\Phi^{s\prime\prime}(\omega)$ of the solutions
for the second correlator of the two--component model defined in
Sec.IVA with $v^{s}$ chosen so that $f^{s}=0.9$. The first
correlator needed as an input for the memory kernel (\ref{Gl26}) is the
one corresponding to the $n=0$ curve in Fig. \ref{fig_F12_deep}.
The dashed lines are the stiff--glass approximations, Eqs. (\ref{Gl28}), 
to these curves. The arrows indicate, which axis corresponds to the
curves.
}
\label{fig_FS_deep}
\end{figure}

\begin{figure}
\caption{Rescaled tagged--particle--density--fluctuation spectra
$10^6 \Phi_q^{s\prime\prime}(\omega) / q^2$ of the HSS for the
packing fraction $\varphi = 0.600$ at wave numbers $q = 3.4$
(solid), $q = 7.0$ (dashed), and $q = 10.6$ (dotted).
}
\label{linsisfw}
\end{figure}

\begin{figure}
\caption{Density of states $g(\omega)$ of a HSS, calculated for
the packing fraction $\varphi=0.60$. The dashed line shows
$g(\omega)= 9.2\times 10^{-8}\omega^{2}$, describing the
density--of--states asymptote at small frequencies. The arrow marks
the Debye frequency $\omega_D=309$ for the longitudinal sound.
}
\label{fig_dos}
\end{figure}

\begin{figure}
\caption{Dynamic structure factor
$S(q,\omega)=S_q\Phi_q^{\prime\prime}(\omega)$ of the HSS for
packing fraction $\varphi=0.60$ as a function of the wave number
$q$ at some fixed frequencies $\omega$. The Debye vector is
$q_D=4.08$; threshold and maximum of the boson peak are located
near $\omega=45$ and $\omega=85$ 
respectively (compare Fig. \ref{fig_scat_cont}).
The lines are guides to the eye.
}
\label{fig_Sqw}
\end{figure}

\end{document}